\documentclass[iop,twocolappendix,numberedappendix,apj]{emulateapj}
\usepackage{epsfig, natbib, graphicx, color, amsmath, cancel,icomma,bm}

\newcommand{\redmapper}{redMaPPer}
\newcommand{\icmod}{i_{\mathrm{cmod}}}
\newcommand{\rcmod}{r_{\mathrm{cmod}}}
\newcommand{\umod}{u_{\mathrm{mod}}}
\newcommand{\gmod}{g_{\mathrm{mod}}}
\newcommand{\rmod}{r_{\mathrm{mod}}}
\newcommand{\imod}{i_{\mathrm{mod}}}
\newcommand{\zmod}{z_{\mathrm{mod}}}
\newcommand{\teff}{t_{\mathrm{eff}}}

\newcommand{\mgal}{m_{\mathrm{gal}}}
\newcommand{\merr}{m_{\mathrm{err}}}
\newcommand{\mlim}{m_{\mathrm{lim}}}

\newcommand{\mzp}{m_{\mathrm{ZP}}}
\newcommand{\sigresid}{\sigma_{\mathrm{resid}}}
\newcommand{\sigsim}{\sigma_{\mathrm{sim}}}

\newcommand{\be}{\begin{equation}}
\newcommand{\ee}{\end{equation}}
\newcommand{\avg}[1]{\langle {#1} \rangle}
\newcommand{\Flim}{F_\mathrm{lim}}
\newcommand{\Fnoise}{F_\mathrm{noise}}
\newcommand{\sigmaobs}{\sigma_m^\mathrm{obs}}

\newcommand{\bn}{\bm{\hat n}}
\newcommand{\bp}{\bm{p}}

\newcommand{\healpix}{HEALPix}

\shortauthors{Rykoff et al.}
\shorttitle{Photometric Survey Depth}

\begin{document}
\title{Assessing Galaxy Limiting Magnitudes in Large Optical Surveys}
\author{E.~S.~Rykoff\altaffilmark{1,2},
  E.~Rozo\altaffilmark{3},
  R.~Keisler\altaffilmark{1}
}
\altaffiltext{1}{Kavli Institute for Particle Astrophysics \& Cosmology, P.
O. Box 2450, Stanford University, Stanford, CA 94305, USA}
\altaffiltext{2}{SLAC National Accelerator Laboratory, Menlo Park, CA
94025, USA}
\altaffiltext{3}{Department of Physics, University of Arizona, Tucson, AZ
85721, USA}

\begin{abstract}
Large scale structure measurements require accurate and precise knowledge of
the survey depth --- typically expressed in the form of a limiting magnitude
--- as a function of position on the sky.  To date, most surveys only compute
the point-source limiting magnitude measured within a fixed metric aperture.
However, this quantity is ill suited to describe the limiting depth of
galaxies, which depends on the detailed interplay of survey systematics with
galaxy shapes and sizes.  We describe an empirical method for directly
estimating the limiting magnitude for large photometric surveys, and apply it
to $\sim10,000\,\mathrm{deg}^{2}$ of SDSS DR8 data.  Combined with deeper
imaging from SDSS Stripe 82 and CFHTLens, we are able to use these depth maps
to estimate the location-dependent galaxy detection completeness at any point
within the full BOSS DR8 survey
region.  We show that these maps can be used to construct random points
suitable for unbiased estimation of correlation functions for galaxies near the
survey limiting magnitude.  Finally, we provide limiting magnitude
maps for galaxies in SDSS DR8 in \healpix{} format with NSIDE=2048.
\end{abstract}


\section{Introduction}

In recent years, observational astronomy has been revolutionized by wide-field
optical surveys.  The impact of the data from the Sloan Digital Sky
Survey~\citep[SDSS][]{yorketal00} has been enormous, covering a range of
topics too broad to mention.  The next era of large photometric surveys is now
upon us, with the Dark Energy
Survey~\citep[DES][]{des05} underway, imaging 5,000\ deg$^2$ with a depth that is 
two magnitudes deeper than that of SDSS.  Similar surveys like the Kilo-Degree
Survey~\citep[KiDS][]{kidsdr1} and the Hyper-Suprime Camera\footnote{http://http://www.naoj.org/Projects/HSC/} (HSC) are also underway.
By the end of the decade the Large Synoptic Survey
Telescope~\citep[LSST][]{lsst08} will start to take data, with unprecedented
depth over fully half the sky.  However, as hard as we try to build homogeneous
surveys with uniform depth, there will always be variations in sky coverage,
with exposure time, seeing, sky brightness, and other factors varying over the
course of nights and seasons.  These systematic variations can impact
the scientific products derived from these surveys, affecting
spectroscopic target selection, optical cluster measurements, and
correlation function measurements (to name but a few).

One obvious option to mitigate these factors is to limit any object selection
to those regions where the imaging is nearly complete.  This of course raise
the question of choosing which regions have high completeness.  One option is
to mask out regions which have anomalously low object
density~\citep[e.g.][]{liu+15}.  Other options include cross-correlating galaxy
densities with known sources of systematic errors in order to choose which
regions were contaminated and should be masked, as for early studies with
SDSS~\citep{sjdfc02}.  The Baryon Oscillation Spectroscopic
Survey~\citep[BOSS][]{schlegel09} made use of relatively faint targets, requiring more
sophisticated analyses~\citep{rhctp11}.  Galaxy detection probabilities were
estimated via cross-correlation with systematics maps, and galaxies used in the
correlation function measurements were reweighted accordingly.  However, these
effects were mostly perturbative for the spectroscopic targets considered.

The WiggleZ spectroscopic survey~\citep{blake10} had a much more challenging
task due to the combination of SDSS and Galaxy Evolution
Explorer\footnote{http://www.galex.caltech.edu} (GALEX) target selection.
Density variations as a function of Galactic dust and GALEX exposure time were
modeled via a simple completeness function.  These models were then used to
compute the full survey selection function. Most recently, a significant amount
of effort has gone into simulating the selection function directly over a large
survey by inserting and measuring fake galaxies with known
properties~\citep[i.e. the Balrog code,][]{suchyta15}.  Although this may be the most robust
method, it suffers from a couple limitations.  First, it can be
expensive\footnote{Though of course we should be willing to pay for expensive
  simulations to match our expensive surveys.}.  Second, inserting fake
galaxies will tend to correlate existing structure with the selection function,
somewhat muddying the interpretation of the resulting random catalogs.   Of course,
this effect is in the real data as well, but since the interpretations are non-trivial,
having a completely independent approach for characterizing selection functions can
be highly complementary.

In this paper, we suggest an alternative, pragmatic approach.
The main idea behind our measurement is simple: the magnitude error $\Delta m$
of a galaxy should be a function of the galaxy's magnitude and the survey
depth: the deeper the survey, the smaller the photometric error $\sigma_m$.
Given a model for magnitude errors as a function of magnitude and survey depth
$\sigma_m = g(m|\mlim)$, we use observations of galaxies $(m,\Delta m)$ over a
patch of sky to recover the limiting magnitude $\mlim$ of the survey patch.  In
practice, this direct measurement of the survey depth can only be done in
relatively coarse pixels.  We overcome this difficulty by fitting the recovered
depth map as a function of the systematics map for the survey (e.g. sky noise,
PSF size, etc), and then utilize these fine structure maps to generate a final
high resolution estimate of the survey depth.

The layout of this paper is as follows.  In Section~\ref{sec:data} we describe
the SDSS data used in this paper, although our method is fully generalizable to
any survey data.  In Section~\ref{sec:depth} we describe our method for
measuring the survey depth in coarse pixels on the sky, and in
Section~\ref{sec:recon} we show how we can use machine learning methods
to reconstruct the survey depth at high resolution by making use of detailed
maps of survey systematics.  In Section~\ref{sec:completeness}, we use deeper
data to model the galaxy detection completeness, and show how this is a simple
function of survey depth.  In Section~\ref{sec:corrfxn} we show a practical
application of our method in a simulated computation of the correlation
function near the survey limiting magnitude where depth and completeness
variations are significant.  Finally, in Section~\ref{sec:summary} we summarize
our findings.


\section{Data}
\label{sec:data}

As discussed above, our method is designed to handle an arbitrary photometric
galaxy catalog.  As a case study, in this paper we determine the depth of SDSS DR8
photometric data.

\subsection{SDSS DR8 Photometry}

The input galaxy catalog for this work is derived from SDSS DR8
data~\citep{dr8}.  This data release includes more than $14,000\,\mathrm{deg}^2$
of drift-scan imaging in the Northern and Southern Galactic caps.  The survey
edge used is the same as that used for BOSS target selection~\citep{dsaaa13}, which reduces the total area to
$\approx 10,500\ \deg^2$ with high-quality observations and a well-defined
contiguous footprint\footnote{We note that we produce our maps over the full
  DR8 footprint, but all of our verification in this paper are restricted to
  the BOSS footprint.}.

The galaxy selection employed is nearly the same as that used in the construction
of the SDSS DR8 \redmapper\ cluster catalog~\citep{rykoffetal14}. 
The primary difference is that we limit the catalog to
$\icmod<22.0$, rather than $\icmod<21$, to ensure that we have galaxies that
are fainter than the $10\sigma$ limiting magnitude over the full survey footprint.
Furthermore, we have \emph{not} corrected for Galactic reddening.  We
then filter all objects with any of the following flags set in the $g$, $r$, or
$i$ bands: {\tt SATUR\_CENTER}, {\tt BRIGHT}, {\tt TOO\_MANY\_PEAKS}, and ({\tt
  NOT BLENDED} {\tt OR} {\tt NODEBLEND}).

In this work, we reconstruct the SDSS DR8 {\tt MODEL\_MAG} limiting magnitudes
in $u$, $g$, $r$, $i$ and $z$, and the composite {\tt CMODEL\_MAG} limiting
magnitudes in $r$ and $i$.  The {\tt MODEL} magnitudes are estimated by fitting
each of an exponential and deVaucouleurs profile to each galaxy in the $r$
band, and applying the better-fit model to each individual band.  As such, {\tt
  MODEL} magnitudes are suitable for galaxy color measurements.  The composite {\tt
  CMODEL} magnitudes are computed as a linear combination of the best-fit
exponential and deVaucouleurs models, and are useful for galaxy total magnitude
measurements.  In this paper, we denote {\tt MODEL} and {\tt CMODEL} magnitudes
with subscripts ``mod'' and ``cmod'' respectively.


\subsection{SDSS Systematics Maps}

The SDSS survey consists of drift-scan imaging of a large number of stripes
that extend along great circles~\citep{yorketal00}.  Because of the CCD layout,
each stripe is scanned at least twice, with six interleaving camera columns.
These two scans may have different seeing and sky noise parameters, yielding
the possibility of a ``cat scratch'' pattern of imaging systematics.  The SDSS
imaging pipeline divides each camera column into ``fields'', which are
rectangular patches of size $~145\, \mathrm{arcmin}^2$.  Due to the interleaving
nature of the scanning survey, there are overlaps between neighboring fields.

For SDSS DR8, to create a uniform single-pass survey and resolve field
overlaps, each field is given a score based on the $r$-band
seeing, the sky brightness in $r$, and an estimate of photometricity of the
night~\citep{dr8}.  Fields with higher scores are determined to be ``primary'', with an
additional percolation step that tries to keep regions as homogeneous as
possible.  However, as shown below there can still be a large variation in
systematics over small scales at the edges of stripes and camera columns.

At the end of this procedure, each point on the sky is associated with a single
primary field.  The primary window function thus determined is available from
the SDSS-III
website\footnote{http://data.sdss3.org/datamodel/files/PHOTO\_RESOLVE}.  To
implement the window function, the sky is broken into many mangle
polygons~\citep{sthh08}, each of which is linked to a primary field.  In this
way each point in the survey can be uniquely associated with a point spread
function (PSF) full width half maximum (FWHM), sky level, and sky noise in all
five bands.  In addition, the observation time can be used to compute the
airmass~(D. Schlegel, private communication).

For the analysis in this paper, we have converted the window function into
\healpix{} maps with NSIDE=2048.  Each pixel has a size of
$2.95\,\mathrm{arcmin}^2$, which is of sufficiently high resolution to capture the fine
scale variations of the systematics maps.  However, for pixel-level precision
then the full mangle polygon description of the window function should be used.


\section{Measuring Survey Depth}
\label{sec:depth}

\subsection{Relating Survey Photometry to Survey Depth}
\label{sec:model}

Let $\mgal$ be the magnitude of a galaxy.  If $F$ is the flux of a
source in units of nanomaggies, the flux is related to galaxy magnitude
via
\be
F = 10^{-0.4(\mgal-\mzp)}.
\label{eq:mag}
\ee
where $\mzp=22.5$.
If $S$ is the number of signal photons reaching the detector, one has that the expectation value for
$S$ is related to the galaxy flux via
\be
\avg{S} = k \teff F
\ee
where $k$ is a proportionality constant, and $\teff$ is the effective exposure time.  
The corresponding number of noise-photons $N$ is
\be
\avg{N} = k \teff \Fnoise
\ee
where $\Fnoise$ is the effective noise flux. The noise flux can have multiple
origins, e.g. sky, read noise, etc.  Here, we have lumped any such contributions into a single term.
Given $N$ signal photons, an unbiased estimate for the galaxy flux is
\be
\hat F = \frac{S+N}{k\teff} - \Fnoise.
\ee
Since both $S$ and $N$ follow Poisson statistics, the variance in $\hat F$ is
\be
\sigma_F^2 = \frac{\avg{S}+\avg{N}}{k^2\teff^2} = \frac{F+\Fnoise}{k\teff}
\ee
The corresponding magnitude error is
\begin{equation}
\begin{split}
 \sigma_m (F;\Fnoise,\teff) &= \frac{2.5}{\ln
   10}\frac{\sigma_F}{F} \label{eq:sm}\\
 &= \frac{2.5}{\ln 10}\left [ \frac{1}{F k\teff} \left ( 1 + \frac{\Fnoise}{F}  \right ) \right ]^{1/2}
\end{split}
\end{equation}

The above expression relates the flux of a galaxy to its corresponding magnitude error.  In practice, we recast
$\Fnoise$ in terms of the $10\sigma$ limiting flux $\Flim$ defined via $\Flim/\sigma_F=10$.  Solving for $\Fnoise$,
we have
\begin{equation}
\Fnoise = \frac{\Flim^2k\teff}{10^2} - \Flim
\label{eq:fn}
\end{equation}
The limiting magnitude $\mlim$ is simply the magnitude associated with a galaxy of flux $\Flim$.  

Equations~\ref{eq:mag}, \ref{eq:sm} and \ref{eq:fn} together define our model for the magnitude error 
$\sigma_m(m;\mlim,\teff)$ of a galaxy of magnitude $m$ in a survey patch of limiting magnitude $\mlim$
and effective exposure time $\teff$.  
Note that in practice, the effective time $\teff$ is always accompanied by the normalization constant $k$, which can
be absorbed into the definition of the parameter $\teff$.

{\bf A note on dust and dereddened magnitudes:} Note that in the above derivation, the relevant quantity is the observed magnitude
$m$, or the observed flux $F$.  Whether the flux has been attenuated due to dust or not is completely irrelevant; the only thing 
that matters is the total arriving flux.   Consequently, one should {\it not} use dereddened
magnitudes when estimating the limiting magnitude of a survey.   

Conversely, we emphasize that the limiting magnitude we recover is a property
of the survey.  To relate our measured limiting magnitude to a dereddened
limiting magnitude, one must apply the necessary reddening corrections.  As
part of the data release associated with this paper, for convenience we include the necessary
dust maps in the equatorial coordinate system that we use.  We have chosen not
to apply these maps to allow for the use of different Galactic dust models.


\subsection{Computing the Depth Maps}
\label{sec:computedepth}

As a practical application, we fit our model to SDSS DR8 data. Specifically,
given a patch of sky, we fit our model of magnitude errors to the reported
magnitude errors so as to derive the effective limiting magnitude and exposure
times.  In this procedure, we assume that the errors as quoted
are correct.  Observationally, while we have magnitude error estimates, we do
not have ``errors on the error'', so it is not obvious how to best fit the
observational data to our model.  In order to minimize the impact of gross
outliers, we have chosen to fit our model by minimizing the total absolute
deviation from the model, i.e. we minimize the cost function
\be
E(\mlim,\teff) = \sum_\alpha \left\vert \sigmaobs  - \sigma_m(m_\alpha|\mlim,\teff) \right\vert
\ee
where the sum is over all galaxies within a given sky patch. 
The function is minimized using the downhill-simplex method of
\citet{nelder65} as implemented in the IDL \texttt{AMOEBA} function, and errors on
the parameters are derived by bootstrap resampling the galaxies 50 times and
refitting.

\begin{figure}
  \begin{center}
    \scalebox{1.0}{\plotone{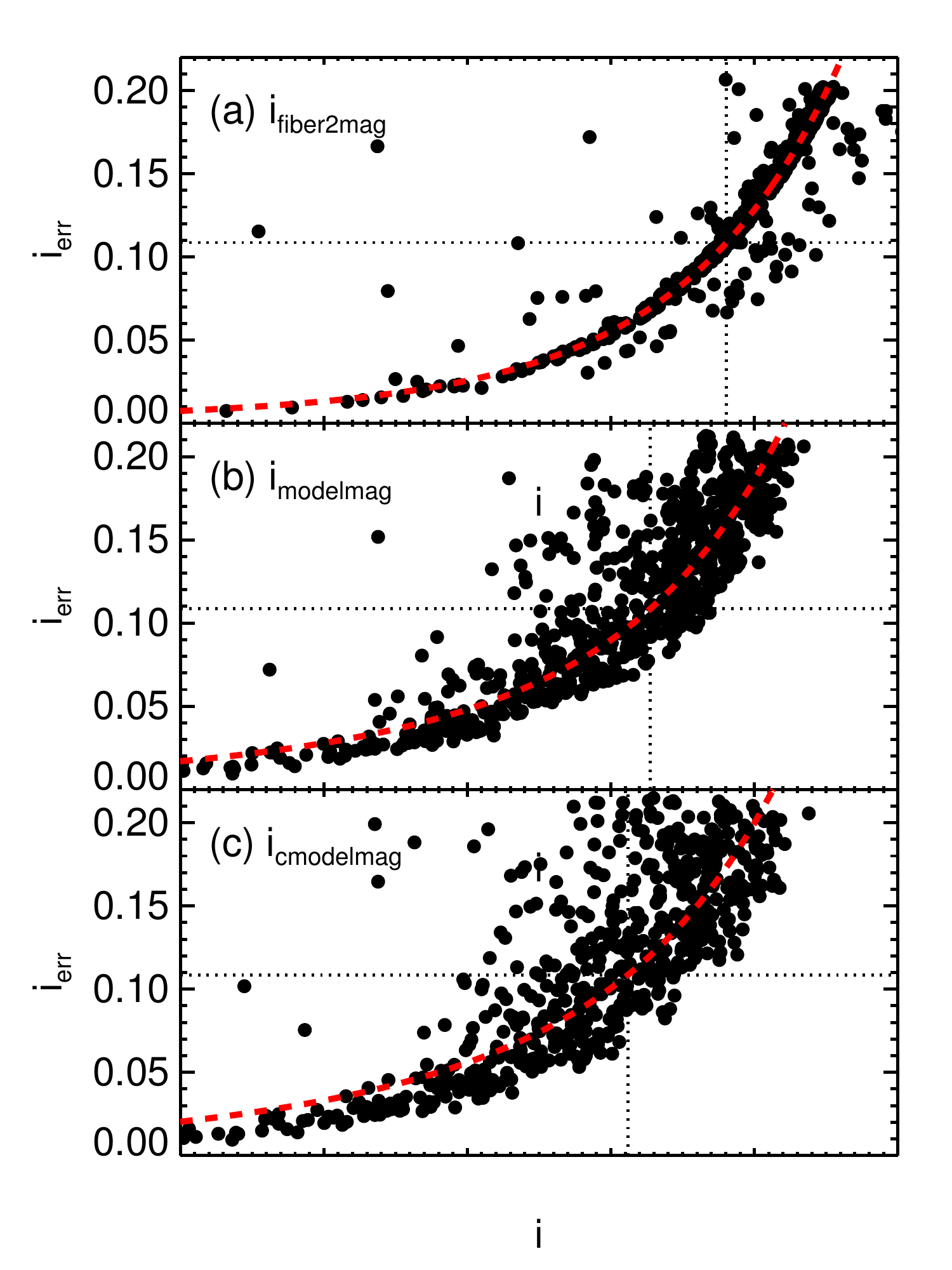}}
    \caption{Magnitude error vs. magnitude for three magnitude definitions in a
      single \healpix{} pixel of NSIDE=256, as follows: \emph{(a)} SDSS {\tt FIBER2MAG}, an
      aperture magnitude of diameter 2\arcsec.     \emph{(b)} Same as above for {\tt MODEL\_MAG}.   
      \emph{(c)} Same as above for {\tt CMODEL\_MAG}.      
      Red dashed lines show our model fits, and
      black dotted lines mark the recovered $10\sigma$
      limiting magnitude. Note our model provides an excellent description of the noise
      for fixed metric aperture magnitudes, while model fitting magnitudes result in noisier
      fits due to variations introduced by galaxy size.  The outlier points in
      the top panel are from regions that have a different local depth in our
      relatively large pixel.}
    \label{fig:depthmod}
  \end{center}
\end{figure}

Figure~\ref{fig:depthmod} illustrates this technique for a single \healpix{}
pixel of NSIDE=256 in DR8.  We fit all galaxies with signal-to-noise (S/N) greater
than $5$ in the pixel, which is deep enough to get a good measurement of the $10\sigma$
magnitude limit, but shallow enough to ensure that the SDSS ``luptitudes''
(arcsinh magnitudes) are equivalent to logarithmic magnitudes.  In
all panels in the figure, red dashed lines are our model fit.   The black dotted lines show 
our recovered $10\sigma$
limiting magnitude.  


The different panels corresponds to different magnitude definitions.  The top
panel uses SDSS {\tt FIBER2MAG}, an aperture magnitude of diameter 2\arcsec.
We see that in this case the model provides an excellent description of the
data, as we would expect.  The outlier points are from regions that have a
different local depth than the bulk of the relatively large pixel.
Section~\ref{sec:recon} demonstrates how we deal with this issue in our
final depth map estimates.  
The middle and lower panels show the $i$-band {\tt MODEL} and {\tt CMODEL} magnitudes,
which are derived from model fits to the galaxy data.  We see that in this case
there is significant scatter due to the fact that the photometric noise for
these model-fit apertures must necessarily depend on additional variables such
as galaxy size.  Nevertheless, it is clear that our model still provides a
reasonable description of the relation between galaxy magnitude and magnitude
error, especially at the faint end where we are trying to estimate the limiting
magnitude.

In order to achieve a reliable fit of $\merr$ vs. $\mgal$, we find we require
at least 100 galaxies with signal-to-noise greater than $5$.  At the typical
limiting magnitude of the SDSS survey, this is achieved by pixelizing the sky
with \healpix{} NSIDE=256, or $190\,\mathrm{arcmin}^2$ per pixel.  Note this is
significantly coarser than the variations of the systematics maps, which
introduces some noise to our fits which we will quantify in
Section~\ref{sec:estnoise}.  
In some cases, especially at the survey edges and for most of the survey in the
much shallower $u$ band, there are insufficient galaxies to perform the fit
with NSIDE=256.  When this happens we expand to the next coarsest pixelization
(NSIDE=128), and if this still does not yield enough galaxies we use NSIDE=64.

\begin{figure*}
  \begin{center}
    \scalebox{1.0}{\plotone{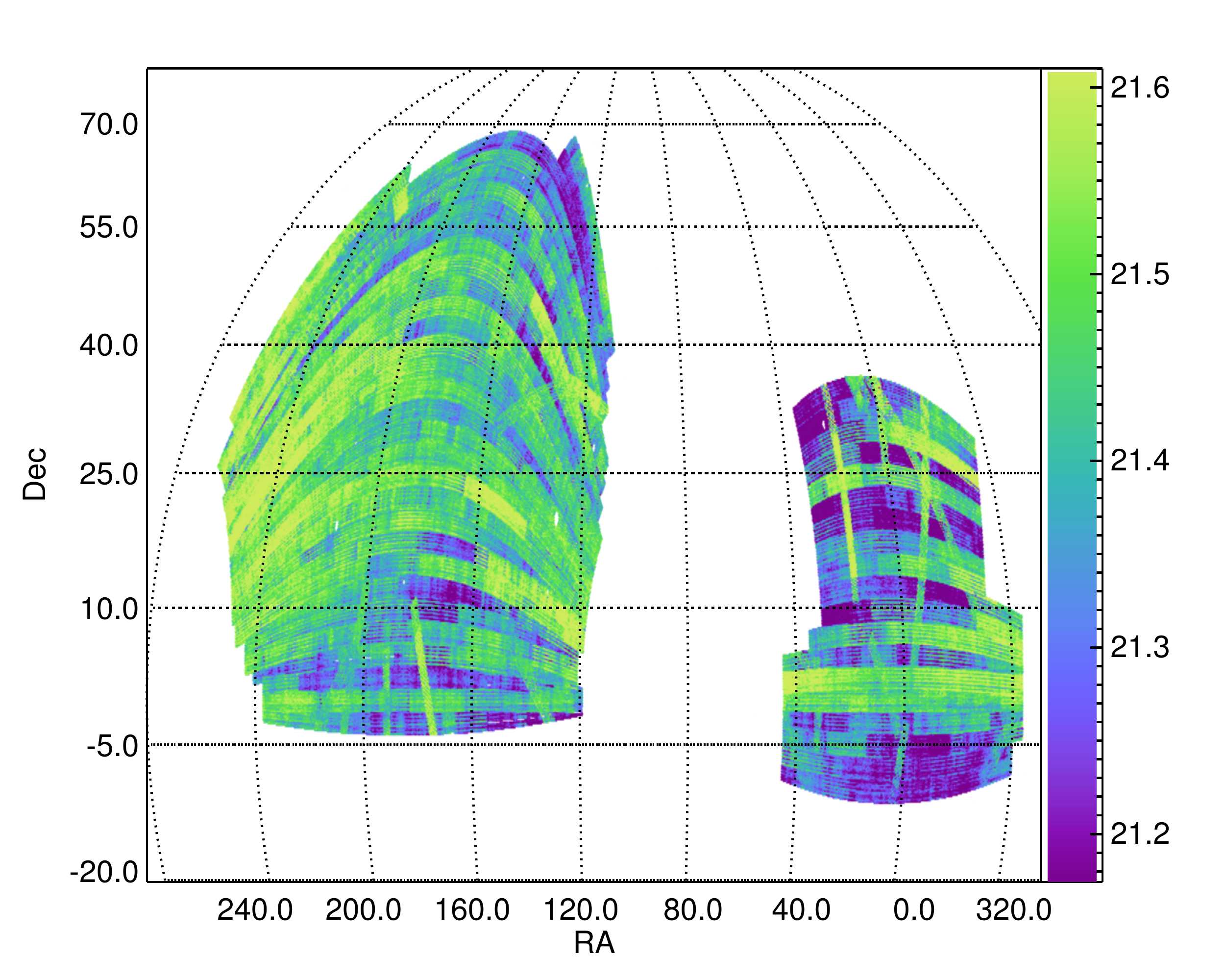}}
    \caption{Depth map for $\rcmod$.  The amplitude of the depth variations is
      $0.5\,\mathrm{mag}$, which will be larger after correcting for Galactic
      reddening.  There are numerous features that are described in the text in
      Section~\ref{sec:computedepth}.}
    \label{fig:depthrcmod}
  \end{center}
\end{figure*}

Figure~\ref{fig:depthrcmod} shows the measured depth map for $\rcmod$.  There
is significant structure on all scales, with an amplitude of $0.5\,\mathrm{mag}$.  
There are several striking features in the map.  First, the stripe
scanning pattern is immediately apparent, as are the ``cat scratch'' features from
interleaving camera columns taken under different conditions.  Second, when the
perpendicular ``\"{u}bercal'' stripes were taken under favorable conditions,
these were chosen as primary fields, and thus are significantly deeper.  Third,
the northern Galactic cap (NGP) region is typically deeper than the
southern Galactic cap (SGP) region. The obvious exception is Stripe 82 in the south.
This region was scanned multiple times, and thus only
the best observations were determined to be primary fields for DR8.


\begin{figure}
  \begin{center}
    \hspace{-0.35in} \scalebox{1.25}{\plotone{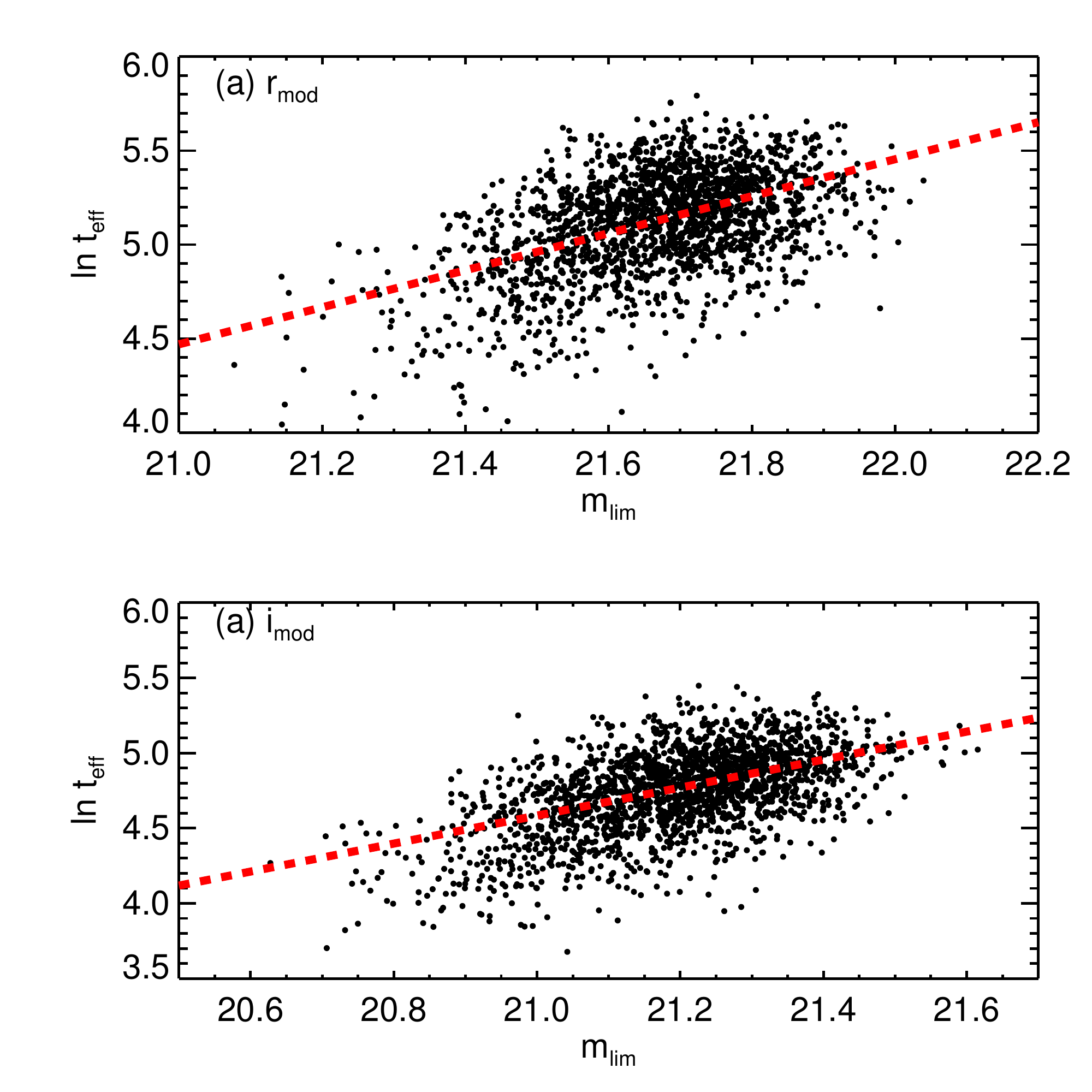}} 
    \caption{$\teff$ vs. $\mlim$ for \emph{(a)} $\rmod$ and \emph{(b)} $\imod$,
      shown for a sample of pixels for clarity.  Red dashed line shows a
      least-absolute-deviation fit.  The residual scatter is $\sim20\%$.}
    \label{fig:teffmlim}
  \end{center}
\end{figure}

After performing the two-parameter fit to each of the pixels independently, we
note that the $\teff$ and $\mlim$ parameters are highly correlated.  This is
illustrated in Figure~\ref{fig:teffmlim} where we show a sample of pixels for
both $\rmod$ and $\imod$.  We fit a linear model
\begin{equation}
  \ln\teff = a + b(\mlim - 21.0),
  \label{eqn:teffmlim}
\end{equation}
where 21.0 is a convenient pivot point.  The residual scatter from the model in
both cases is $\sim20\%$.


In light of the correlation between limiting magnitude and survey depth, we
have chosen to characterize the survey depth with the single quantity $\mlim$
(the more relevant quantity), setting the effective exposure time via equation
\ref{eqn:teffmlim}.  Table~\ref{tab:teffmlim} shows the fit parameters for the
magnitudes used in this paper.  The efficacy of this approximation is tested in
the next section.

\begin{deluxetable}{ccc}
\tablewidth{200pt}
\tablecaption{$\teff$ vs. $\mlim$ Fit Parameters}
\tablehead{
  \colhead{Magnitude} &
  \colhead{$a$} &
  \colhead{$b$}
}
\startdata
$\umod$ & 3.41 & 1.15 \\
$\gmod$ & 4.27 & 0.85 \\
$\rmod$ & 4.53 & 0.91 \\
$\imod$ & 4.56 & 1.00 \\
$\zmod$ & 4.39 & 1.34 \\
$\rcmod$ & 4.34 & 1.70 \\
$\icmod$ & 4.48 & 1.23 \\
\enddata
\tablecomments{$\teff = \exp[a + b(\mlim - 21.0)]$}
\label{tab:teffmlim}
\end{deluxetable}


\section{Reconstruction of the Depth Maps}
\label{sec:recon}

\subsection{Introduction to the Problem}

In principle, Figure~\ref{fig:depthrcmod} is exactly what we are looking for: a map of the limiting
magnitude of the survey everywhere in the sky.  In practice, however, our measurement is severely
limited by the need to have $\sim 100$ galaxies in a sky pixel in order to perform our measurement.
This limitation renders the pixels so wide that important features in the small scale structure of the
survey are missed.  

To overcome this difficulty, we rely on the fact that the observed depth map is a natural consequence of
the survey observing conditions.  For instance, better seeing conditions leads to deeper data.  The fundamental
insight is that the limiting magnitude $\mlim(\bn)$ at a position angle $\bn$ is a function of systematic parameters
such as sky noise, seeing, etc.  Let then $\bp(\bn)$ be the vector that collects all systematics parameters 
at position $\bn$.  We posit that the function $\mlim(\bn)$ takes the form
\be
\mlim(\bn) = f( \bp(\bn) )
\ee
where $f$ is some unknown function.  We seek to estimate the function $f$ given a series of 
inputs $\bp(\bn)$ --- the systematics evaluated for each of our coarse pixels --- and a series of outputs
$\mlim$ --- our measured depth.  This is a problem that is well suited to machine learning methods.
Here, we rely on the Random Forest (RF)
technique~\citep{breiman01}, as implemented in the python module {\tt
  sklearn}\footnote{http://scikit-learn.org}.


\subsection{Method}
\label{sec:avgmaps}

The first step is to decide which set of systematics maps will be used to
train the RF method.  We have opted for an inclusive approach, and
attempted to utilize all the systematic information that
may be relevant.  This includes the PSF FWHM, sky level, and sky sigma for
the band being fit; the airmass; and the value of the $E(B-V)$ map from
\citet{sfd98}.  Although we do not expect the depth to depend on $E(B-V)$
directly, there may be some second-order correlations.  As the
model magnitude depends on the galaxy model as fit in the $r$-band, we
additionally include the $r$-band systematics maps (PSF FWHM, sky, and sky
sigma) in each of the fits.  Furthermore, the drift-scan method of SDSS has all
the bands observed at roughly the same time, so the seeing and sky in the
various bands are correlated and thus we can beat down the noise by using
measurements from different bands.  For this reason, when we fit the $r$-band
depth we additionally include the $i$-band systematics maps in the fit.

When training the RF, it is critical that the systematics information be appropriately
handled.  Specifically, because our observed depth is measured over coarse angular
pixels, the high resolution systematics maps must be averaged to the
same coarser resolution employed in our measurements prior to training the
RF.  To perform this ``de-resolution'' we start with the coarse target
NSIDE=256 map.  In the HEALPix nested scheme it is then trivial to find all the
subpixels at NSIDE=2048 for each of the coarse pixels.  We then take the
average of the systematics value in the subpixels that are within the SDSS
footprint, as well as tracking the fraction of subpixels that are outside the
footprint.

Having averaged our systematics map, we train the RF.
In each run, we use half the pixels for training and half for 
validation.  In all cases, we only use coarse pixels that have at least
80\% of the subpixels in the footprint to avoid problems at the
boundaries.  We have also run checks where we train on points co-located on the
sky to ensure that the classifier is not unintentionally picking up on position
information encoded in nearby correlated pixels, and have achieved similar
performance.  However, in the case where we train on the NGP region and apply
to the SGP region (and vice versa) the resulting performance is degraded.
This hints at some subtle differences in the systematics properties
of the north and south of the SDSS region.

Finally, having trained the RF to calibrate the function $\mlim=f(\bp)$ using
our low resolution systematics map, we apply the RF on the high resolution maps
so as to recover a high resolution map of the limiting magnitude $\mlim(\bn)$.


\subsection{Results}
\label{sec:results}

\begin{figure*}
\begin{center}
  \scalebox{1.0}{\plotone{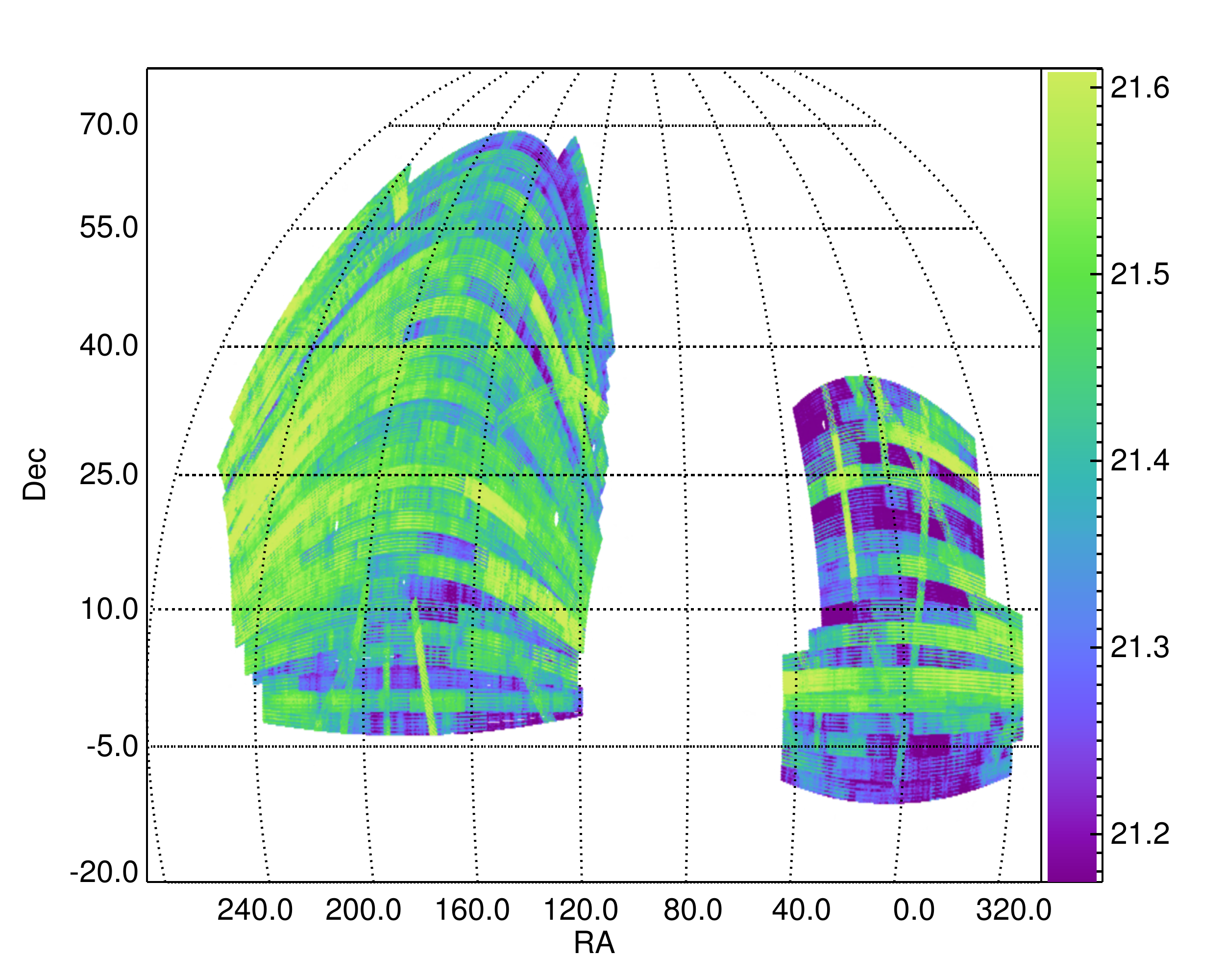}}
  \caption{Reconstructed map of $\rcmod$ depth using RF.  The map was
    reconstructed at high resolution (\healpix{} NSIDE=2048) and then averaged
    to NSIDE=256 as described in Section~\ref{sec:avgmaps} for plotting and
    comparison to the measured map.}
  \label{fig:reconrcmod}
\end{center}
\end{figure*}

\begin{figure*}
  \begin{center}
    \scalebox{1.0}{\plottwo{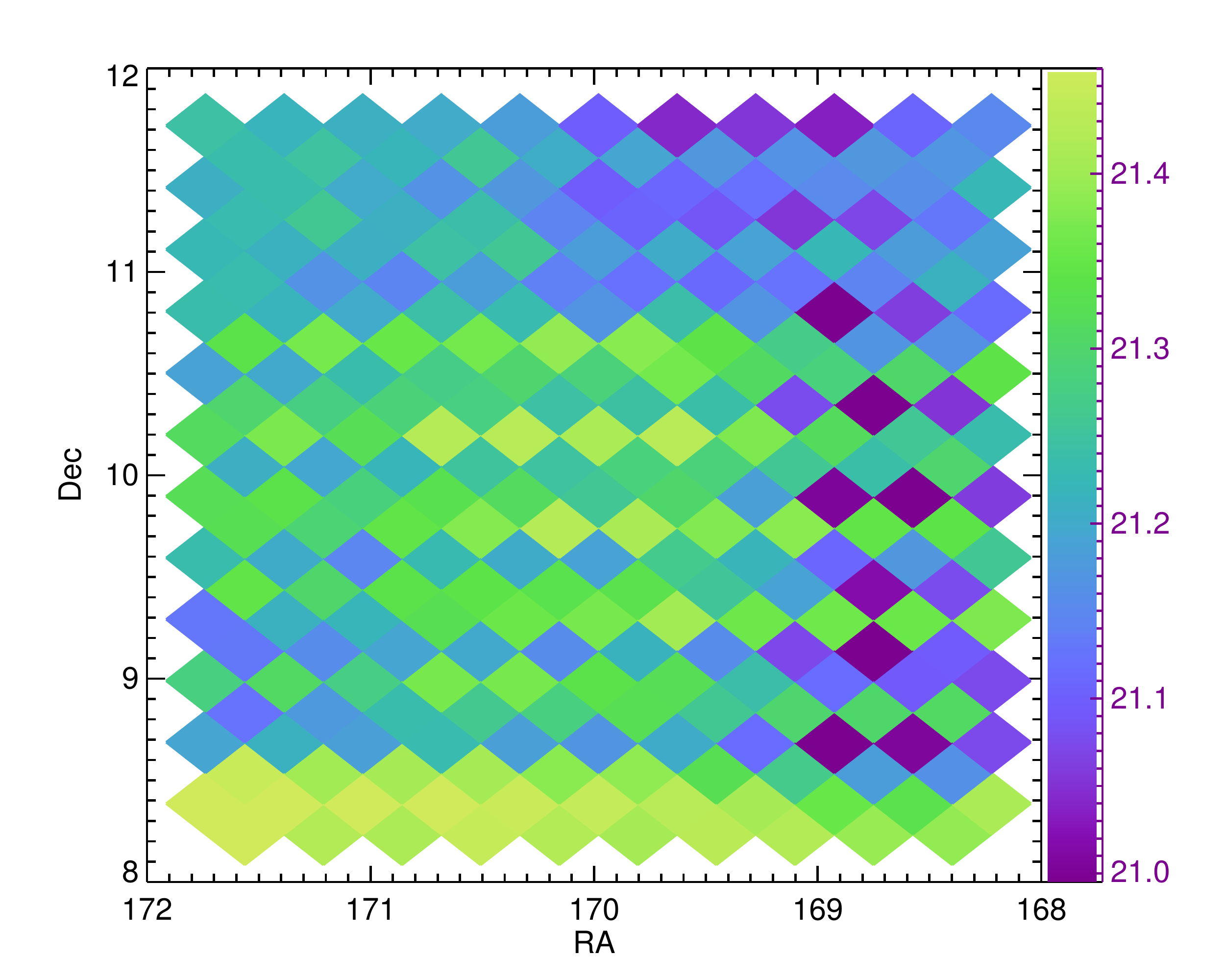}{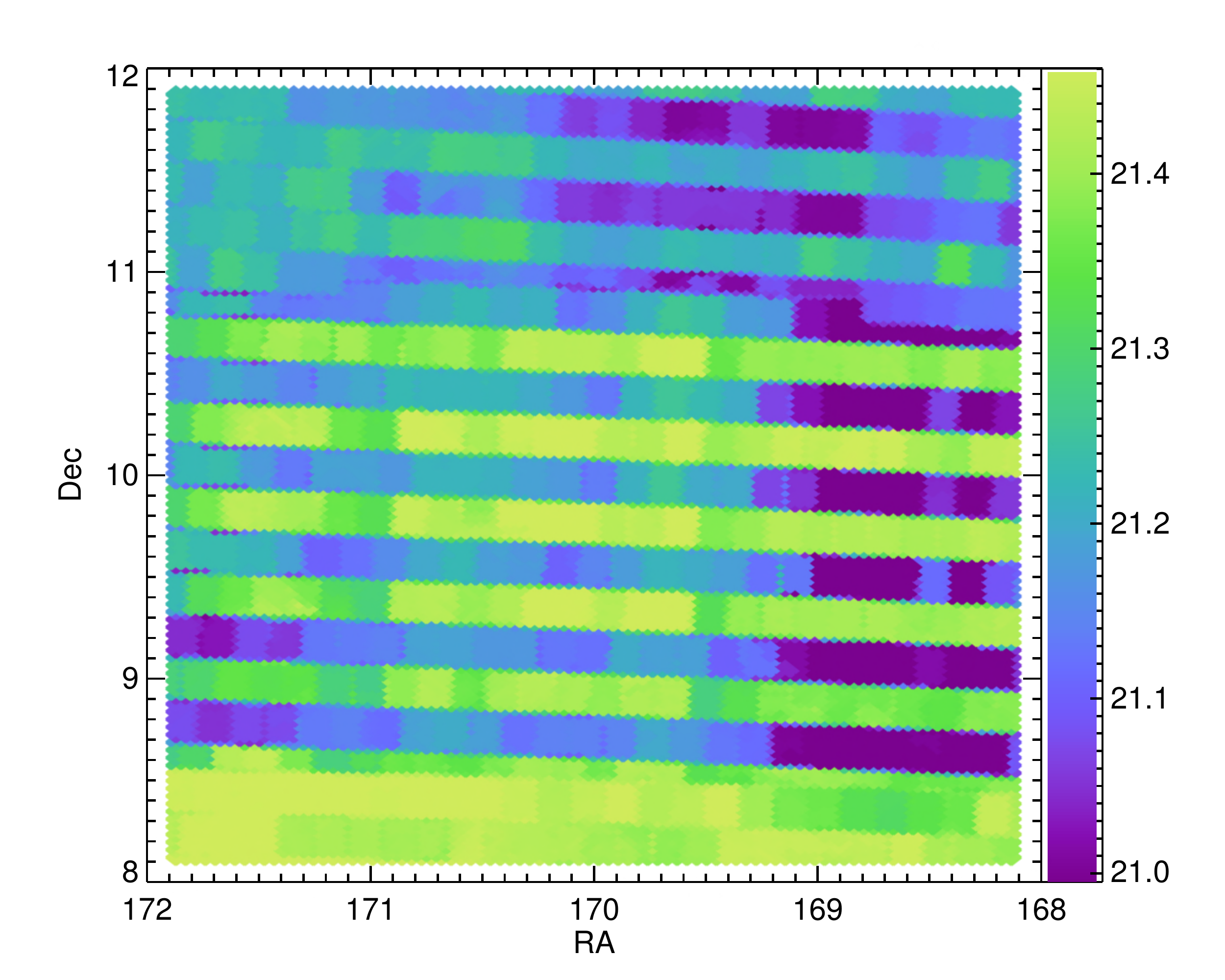}}
    \caption{Zoom in of $16\,\mathrm{deg}^2$ of depth map.  \emph{Left:} Low
      resolution (NSIDE=256) with $\sim200$ galaxies per pixel used for the
      depth estimation.  There is a noticeable gradient, but no higher order
      structure is visible.  \emph{Right:} High resolution (NSIDE=2048)
      reconstructed depth map showing the ``cat scratches'' characteristic of
      the interleaved scanning. These depth variations are tracked by the high
      resolution systematics maps, and are recovered by our reconstruction
      procedure.}
    \label{fig:zoom}
  \end{center}
\end{figure*}

Figure~\ref{fig:reconrcmod} illustrates the fundamental result from our
analysis: a high resolution reconstructed map of the limiting magnitude
$\mlim(\bn)$.  For plotting purposes, the map was averaged to NSIDE=256.  The
structure of the map is very similar to the measured map in
Figure~\ref{fig:depthrcmod}.  Figure~\ref{fig:zoom} highlights the importance
of our reconstruction procedure.  The left panel shows a zoomed-in patch of
$16\,\mathrm{deg}^2$ for the low resolution depth map with $\sim200$ galaxies
per pixel.  While there is a noticeable gradient across this patch, no higher
order structure is visible.  The right panel shows the high resolution
reconstructed depth map for the same patch.  This exhibits the characteristic
``cat scratch'' structure inherent to the interleaved scanning utilized in
SDSS.  These depth variations are tracked by the high resolution systematics
maps, and are recovered by our reconstruction procedure.

\begin{figure*}
\begin{center}
  \scalebox{1.0}{\plotone{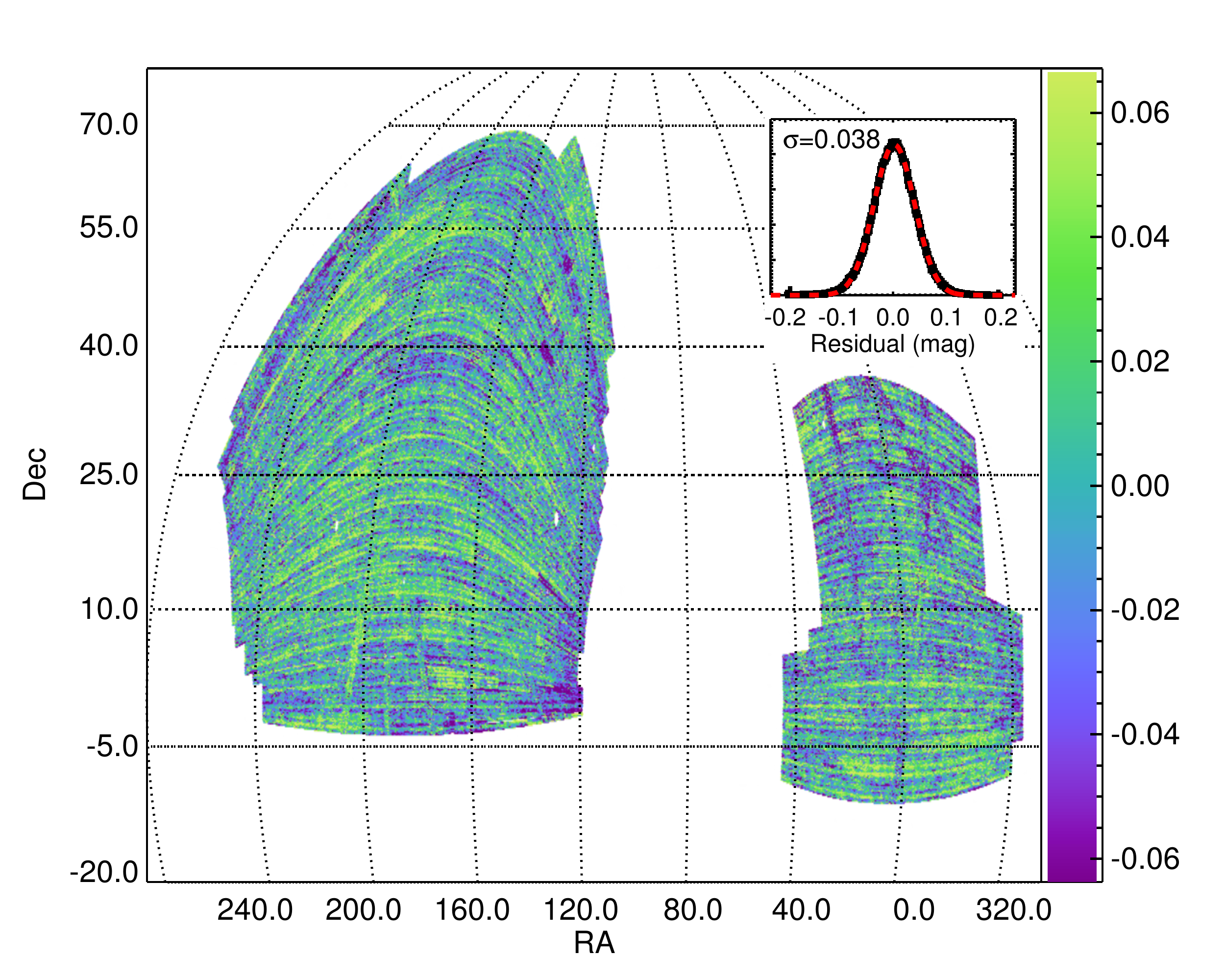}}
  \caption{Residuals of reconstructed $\rcmod$ depth compared to the observed
    limiting magnitude map.  The histogram in the inset plot shows the RMS of
    the residuals, with a Gaussian fit (red) with
    $\sigresid=0.038\,\mathrm{mag}$.  There is some structure in the residuals
    that are visible at a low level.  As shown in Section~\ref{sec:estnoise},
    the large outliers are caused by the averaging process used to compare the
    maps, but are not present in the underlying high resolution map.}
  \label{fig:residrcmod}
\end{center}
\end{figure*}

Figure~\ref{fig:residrcmod} shows the residual between our observed limiting
magnitude map, i.e., Figure~\ref{fig:depthrcmod}, and our low resolution
reconstructed map.  As shown in the histogram in the inset plot, the RMS of the
residuals is $\sigresid=0.038\,\mathrm{mag}$ (for details on how this was
computed, see Section~\ref{sec:estnoise}).  However, there is still some
structure in the residuals that are visible at a low level.  As we will show in
Section~\ref{sec:estnoise}, much of this residual is an artifact of the
interplay of rapidly varying systematics maps and our averaging procedure.

\subsection{Estimating Noise}
\label{sec:estnoise}

We must now estimate the noise in the reconstructed depth maps. The most
obvious estimate is the width of the residual distribution between the measured
depth map and the reconstructed map as plotted in Figure~\ref{fig:residrcmod}.
However, this estimate suffers from some limitations.  Most importantly, our
depth map has been measured at a relatively coarse scale (NSIDE=256), while the
aim of our procedure is to reconstruct the map at a fine scale (NSIDE=2048).
As the averaging procedure adds noise, this will inevitably overestimate the
noise in the residual map as well as introduce spurious structures at field
boundaries where the systematics vary rapidly.

As an alternative, we use a Monte Carlo technique to simulate data with known
systematics and repeat our entire analysis procedure.  A comparison of the
true survey depth map used in the Monte Carlo to the recovered survey depth map 
will allow us to estimate the noise in the reconstructed maps.
For our Monte Carlo simulation, we define the survey depth to be
our estimated high resolution survey depth map, and for our input
galaxy catalog we use the SDSS galaxy catalog, defining the observed
magnitudes as truth for the Monte Carlo simulation.
Our procedure is as follows:
\begin{enumerate}
  \item{Take the existing galaxy catalog, and set each (observed) magnitude as the
    ``true'' magnitude in each band.}
  \item{Take the high resolution reconstructed depth map for each band as the
    ``true'' limiting magnitude (and associated $\teff$) at every point in the
    survey.}
  \item{Given the true magnitude, $\mlim$, and $\teff$, compute the associated
    magnitude error from Eqn.~\ref{eq:sm}.  Perturb the true magnitude
    according to this error.}
  \item{Compute the depth map with NSIDE=256 as described in
    Section~\ref{sec:computedepth}.}
  \item{Refit the depth map and reconstruct a \emph{simulated} high resolution
    depth map as described in Section~\ref{sec:recon}.}
\end{enumerate}
In the end, we have a high resolution reconstructed depth map which can be
directly compared to the ``true'' input map.

\begin{figure*}
  \begin{center}
    \scalebox{1.0}{\plottwo{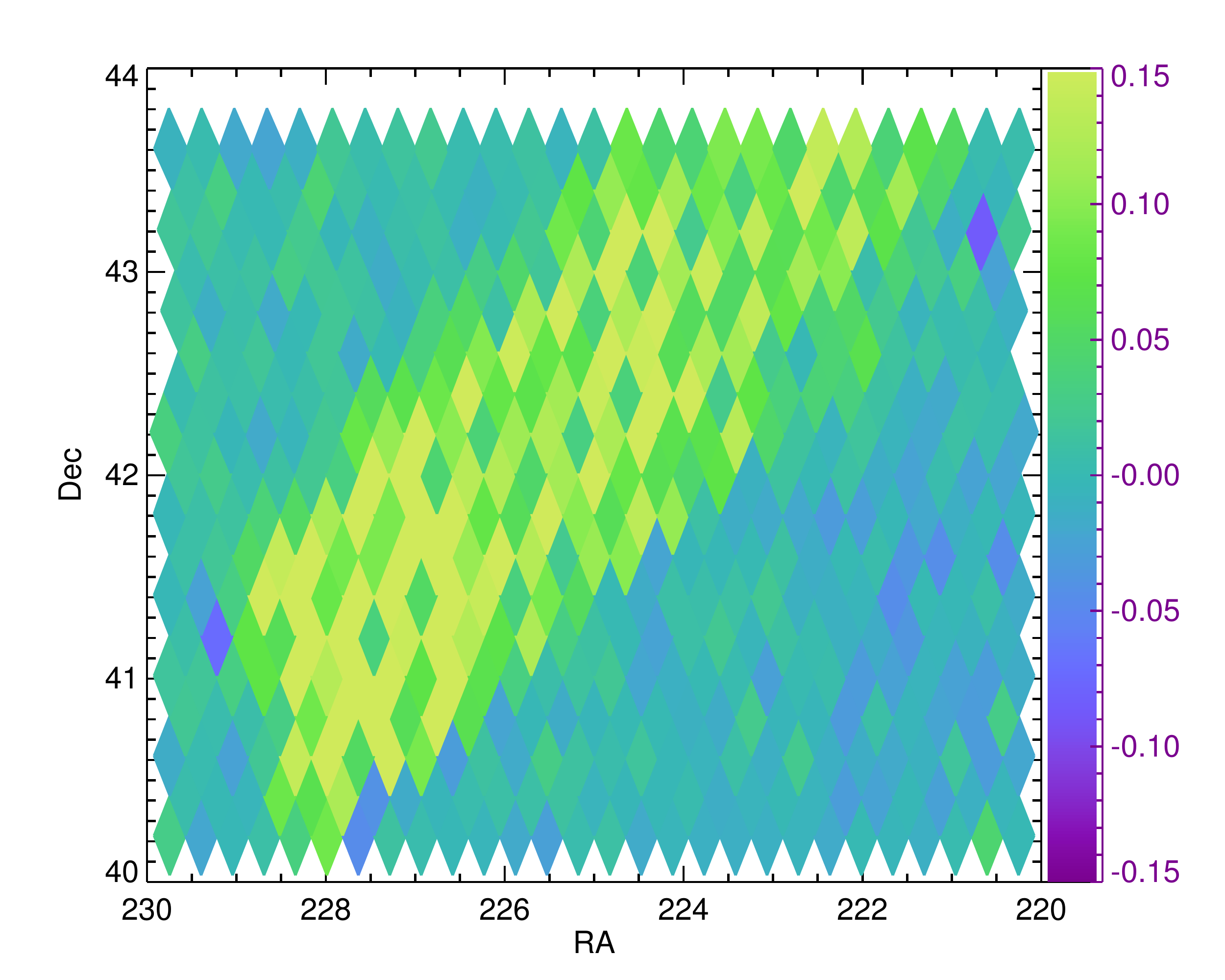}{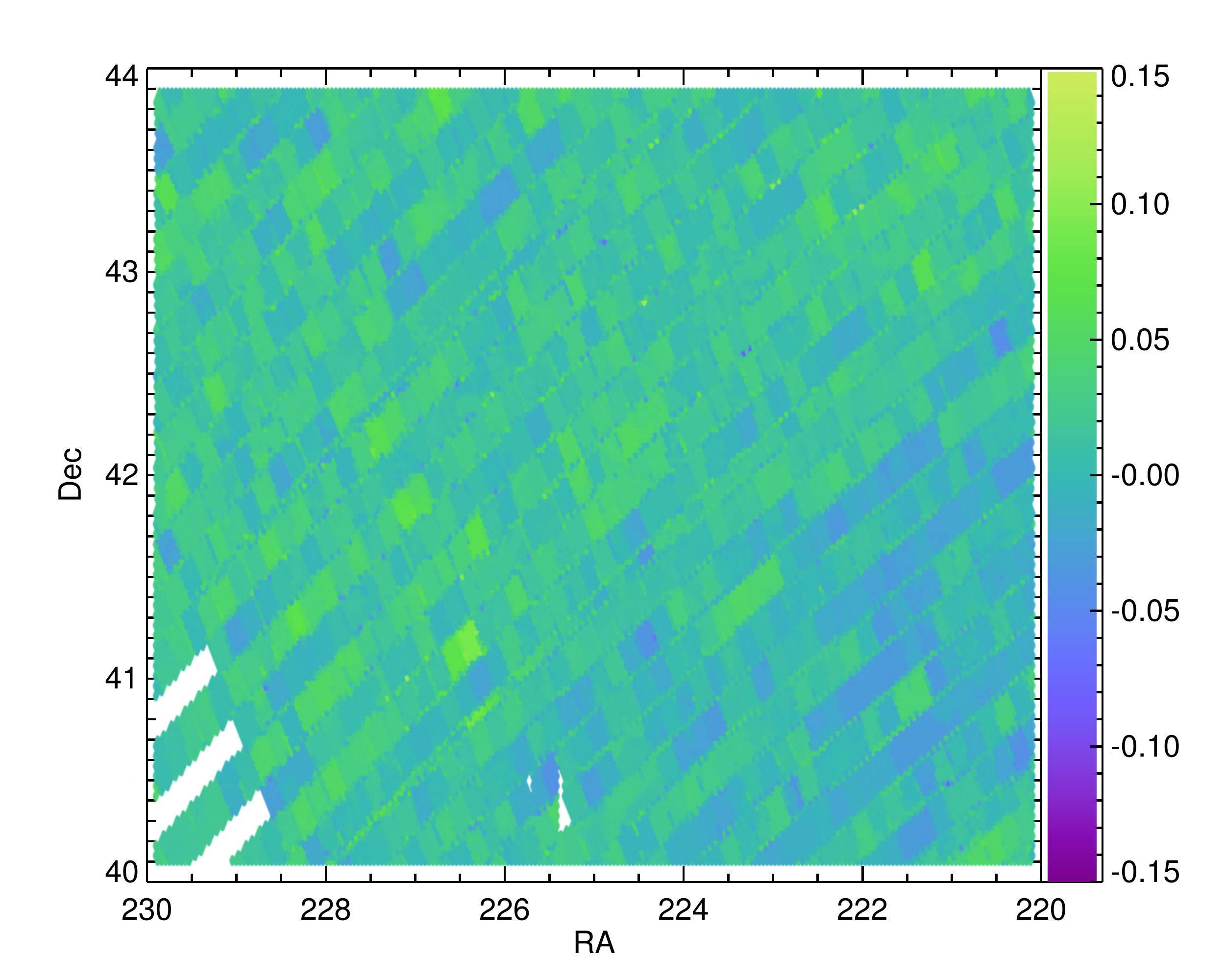}}
    \caption{Zoom in of simulated depth map residuals.  \emph{Left:} Low
      resolution (NSIDE=256) residuals between measured depth map and averaged
      reconstructed depth map for the simulated data.  The red streak shows a
      large offset, similar to that of Figure~\ref{fig:residrcmod}.
      \emph{Right:} High resolution (NSIDE=2048) residuals.  The large offset
      is no longer present, highlighting that our procedure is recovering the
      limiting magnitude at fine scales.}
    \label{fig:reconzoom}
  \end{center}
\end{figure*}

We find that the residual map between the raw measured depth map and our recovered high
resolution depth map (degraded to the resolution of the measured map) 
has a striking similarity to that of Figure~\ref{fig:residrcmod}.
Since we now have the
high resolution ``truth'' in hand, we can determine the origin of the structures shown in Figure~\ref{fig:residrcmod}.
Figure~\ref{fig:reconzoom} shows a particularly
problematic patch at $RA=225^\circ$, $DEC=42^\circ$.   The left panel
shows the NSIDE=256 residuals between the measured depth map and the averaged
reconstructed map for the simulated data.  The red streak shows a large offset
of $>0.1\,\mathrm{mag}$, very similar to that observed in
Figure~\ref{fig:residrcmod} for the real data.  The right panel shows the
NSIDE=2048 high resolution residuals between the reconstructed simulated depth
map and the ``true'' map used as an input.   We see that the pattern essentially
disappears.  Evidently, the regions with large coherent residuals observed
in Figure~\ref{fig:residrcmod} are artifacts of the averaging process.
Most importantly, our high resolution maps are successfully recovering high resolution
features that are completely unresolved in the observed depth maps.

In Table~\ref{tab:obscat}, we summarize the noise estimates for the
reconstructed depth maps for two different methods.  The first, $\sigresid$,
measures the width of the residual distribution between the measured depth map
and the averaged reconstructed depth map (NSIDE=256), as in the inset plot of
Figure~\ref{fig:residrcmod}.  The second, $\sigsim$, is the width of the
residual distribution between the simulated reconstructed depth map and the
``true'' depth map.  In all cases the width is estimated using the robust
Gaussian histogram-fitting code {\tt histogauss.pro} from the IDL Astronomy
Library\footnote{http://idlastro.gsfc.nasa.gov/}.  Due to the fact that our
simulated data cannot take into account all the various sources of unmodeled systematics, we
consider $\sigsim$ and $\sigresid$ to be lower and upper bounds on the true
noise of the reconstructed depth map.

\begin{deluxetable}{ccc}
\tablewidth{0pt}
\tablecaption{Observed and Simulated Scatter}
\tablehead{
  \colhead{Magnitude} &
  \colhead{$\sigresid$} &
  \colhead{$\sigsim$}
}
\startdata
$\umod$ & 0.128 & 0.068 \\
$\gmod$ & 0.038 & 0.020 \\
$\rmod$ & 0.036 & 0.018 \\
$\imod$ & 0.040 & 0.020 \\
$\zmod$ & 0.063 & 0.034 \\
$\rcmod$ & 0.038 & 0.018 \\
$\icmod$ & 0.051 & 0.023 \\
\enddata
\label{tab:obscat}
\end{deluxetable}

\subsection{Using $5\sigma$ Point-Source Depth}
\label{sec:psfmag}

Many sky surveys state depth in terms of $5\sigma$ point-source depth, and this
is typically used for telescope exposure-time calculators.  This has the
advantage of being relatively straightforward to compute, but as we show in
this section this does not scale trivially to the \emph{galaxy} limiting
magnitude which we are interested in computing.

The BOSS survey have independently modeled the $5\sigma$ point-source depth
based on the formal errors from PSF photometry on stellar sources~(D. Schlegel,
private communication).  BOSS uses a simple model of seeing, sky brightness, and
airmass:
\begin{equation}
  F(5\sigma) = a_i f_i \sqrt{s_i} 10^{0.4 k_i m_i},
  \label{eqn:boss5sig}
\end{equation}
where $f_i$ is the FWHM, $s_i$ is the sky flux, and $m_i$ is the airmass.
The normalization term $a = \{0.387, 0.218, 0.241, 0.297, 0.665\}$ and the
airmass term $k = \{0.49, 0.17, 0.10, 0.06, 0.06\}$ for bands $i =
\{u,g,r,i,z\}$ respectively.  This formula predicts the $5\sigma$
point-source depth to an accuracy of 12\%, 4\%, 4\%, 5\%, and 7\%, nominally
comparable to the accuracy of the galaxy depth reconstruction in this work, but
for a different type of object.

\begin{figure*}
  \begin{center}
    \scalebox{1.0}{\plotone{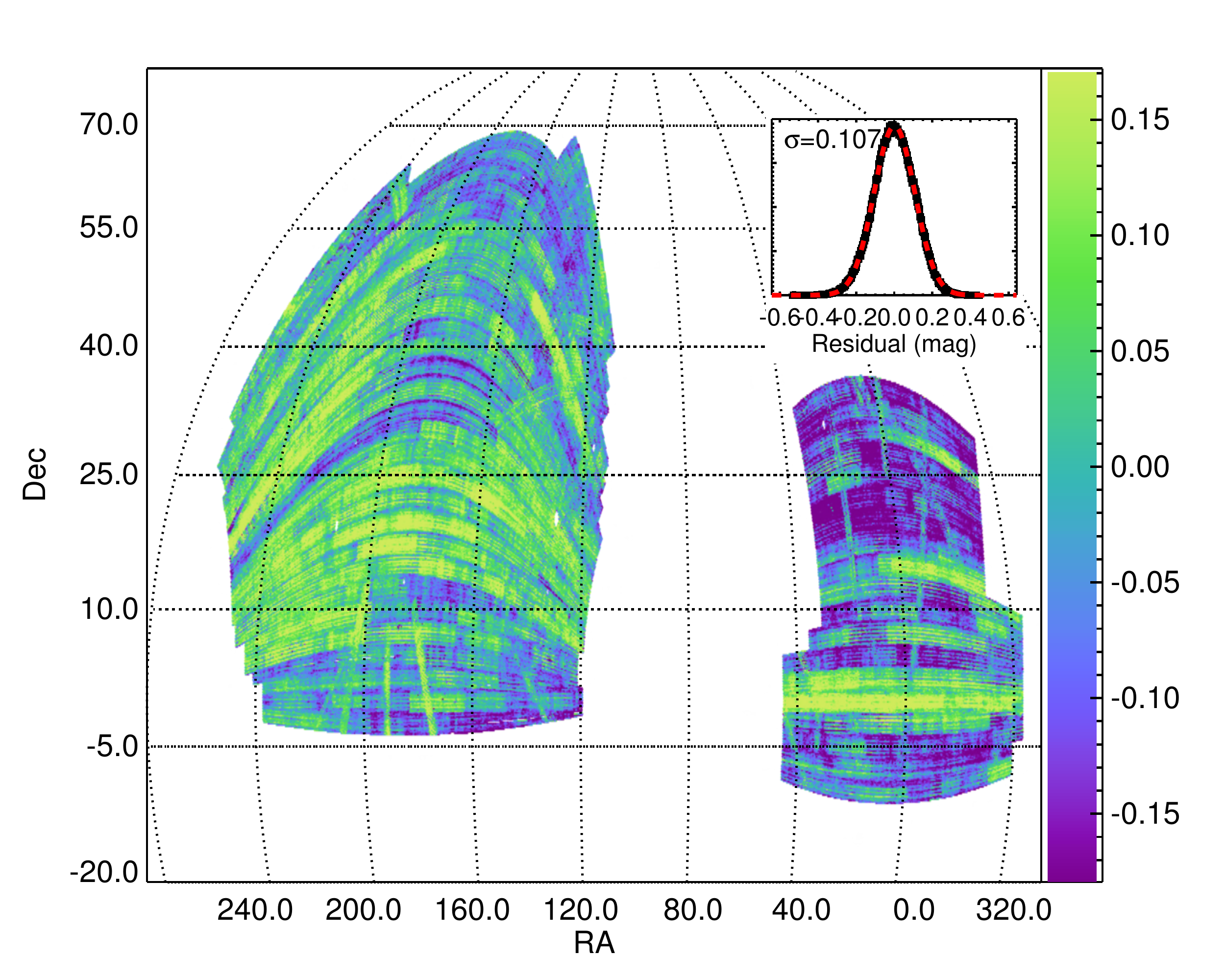}}
    \caption{Residuals of reconstructed $\rcmod$ depth using the formula for
      the $5\sigma$ point-source depth, compared to the measured depth map.
      The RMS scatter is $11\%$, a factor of three larger than seen for our galaxy
      depth reconstruction technique in this paper.  More importantly, there is
      significant structure in the residuals.}
    \label{fig:residrcmodschlegel}
  \end{center}
\end{figure*}

To compare to the galaxy depth, we calculate the $5\sigma$ point-source depth
at high resolution and then average the map to NSIDE=256 as above.  We then
apply a constant offset to account for the fact that the $10\sigma$ galaxy
depth is considerably shallower than that for these point sources.  In the case
of $\rcmod$, this offset is $1.43\,\mathrm{mag}$.
Figure~\ref{fig:residrcmodschlegel} shows the residual between our observed
limiting magnitude map (Figure~\ref{fig:depthrcmod}) and the low resolution map
reconstructed from the point source depth.  The RMS scatter is $11\%$, a factor
of three larger than seen for our galaxy depth reconstruction technique in this
paper.  More importantly, there is significant structure in the residuals.  In
Section~\ref{sec:corrfxn} we show the impact on correlation function
measurements of using this map instead of a galaxy-appropriate map.

\section{Galaxy Catalog Completeness}
\label{sec:completeness}

With our depth maps in hand, we can now properly estimate the completeness of
the galaxy detection as a function of local depth.  We make this estimation
empirically, using two overlapping surveys that are significantly deeper than
single-pass SDSS DR8 imaging.  The first is the SDSS Stripe 82 (``S82'') coadd
catalog~\citep{assbd11}, which covers $\sim 205\,\mathrm{deg}^2$ overlapping our
DR8 footprint, with limiting magnitudes $\sim 2\,\mathrm{mag}$ deeper than the
single-pass imaging.  The second is the CFHTLens catalog~\citep{hvmeh12,ehmvh12}, whose W1 and W4 fields cover $\sim 75\,\mathrm{deg}^2$, with
limiting magnitudes $\sim 3\,\mathrm{mag}$ deeper than DR8.  Along with being
based on two different surveys, these catalogs sample different parts of sky
and different systematics, and as such they are an adequate independent test of
the detection completeness.

\subsection{Measuring Completeness}

We investigate how survey completeness depends on survey depth.
Our method for measuring the completeness of galaxy detection is quite
straightforward.  We start with the ``truth,'' which is the designation we give
to the deeper reference catalog (either S82 or CFHTLens).  Our procedure
is then:
\begin{enumerate}
  \item{Apply the BOSS mask to both surveys.}
  \item{Pixelize the footprint with \healpix{} using NSIDE=2048.}
  \item{Bin the pixels according to local depth, with 10
    bins over S82 and 5 bins over CFHTLens (because of the smaller area).  Note that
    these pixels need not be contiguous; rather, they must share a common local
    depth.}
  \item{For each depth bin, compute a median zero-point correction to put galaxies on
    a ``DR8'' scale.  This removes any overall shifts due to filter and
    magnitude measurement method.}
   \item{Measure the detection rate of galaxies as a
    function of true magnitude in each depth bin.  This is the desired completeness.}   
  \item{Bootstrap resample the pixels in each depth bin 1000 times, remeasuring the depth
  each time in order to determine observational uncertainties.}
  \item{Fit for the completeness as a function of magnitude using the functional form detailed below.  The bootstrap
    resampling of the input pixels is used to estimate errors on parameters.}
\end{enumerate}

\begin{figure}
  \begin{center}
    \scalebox{1.0}{\plotone{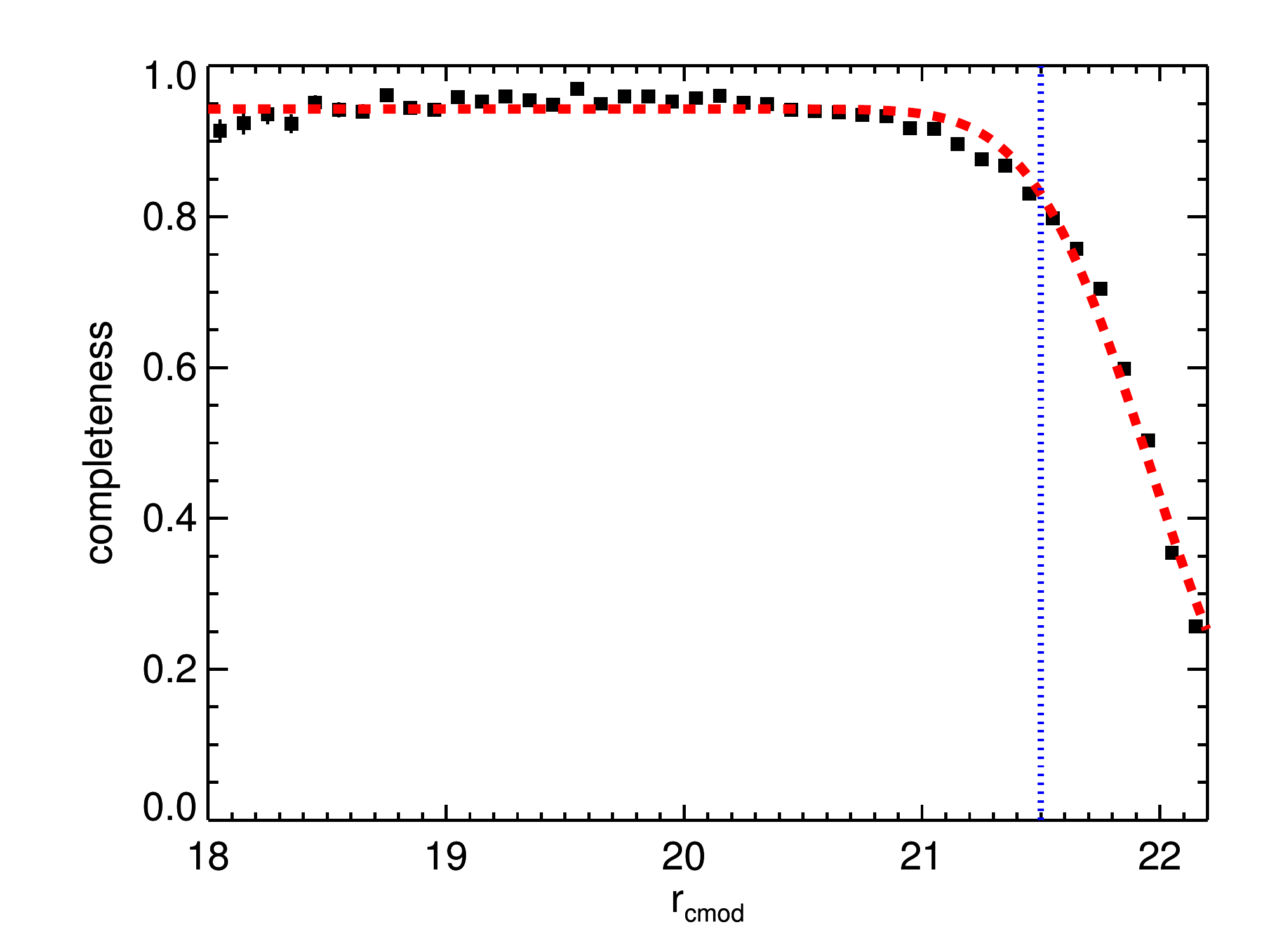}}
    \caption{Completeness estimated for $\rcmod$ for Stripe 82 data for the
      median DR8 depth in the Stripe 82 fields.  Red dashed line shows the fit
      of the functional form given by Eqn.~\ref{eqn:comp}.  The blue dotted
      line shows the $10\sigma$ limiting magnitude.  The error bars are smaller
    than the data points at all but the brightest magnitudes.}
    \label{fig:compvsr}
  \end{center}
\end{figure}

Figure~\ref{fig:compvsr} shows the completeness estimated for $\rcmod$
detections for the median DR8 depth in the S82 fields.  We model the
completeness with a simple error function model:
\be
c = (e/2) \left [ 1 - \mathrm{erf} \left ( \frac{m - m_{50}}{\sqrt{2w}} \right ) \right
],
\label{eqn:comp}
\ee
where $e$ is the overall efficiency of detection at the bright end, $m_{50}$ is
the magnitude at which the completeness is $50\%$, and $w$ is the (Gaussian)
width of the rollover.  The red dashed line in the plot shows the best-fit
model, and the blue dotted line shows the median $10\sigma$ limit in the
sampled region.  The error bars are smaller than the data points at all but the
brightest magnitudes.  Although the model fit is not perfect, it clearly provides
a reasonable description of the shape of the completeness function.
The end result is that we now have measurements of $m_{50}$ and $w$
in bins of different survey depth $\mlim$.

\subsection{Completeness Results}

\begin{figure}
  \begin{center}
    \scalebox{1.0}{\plotone{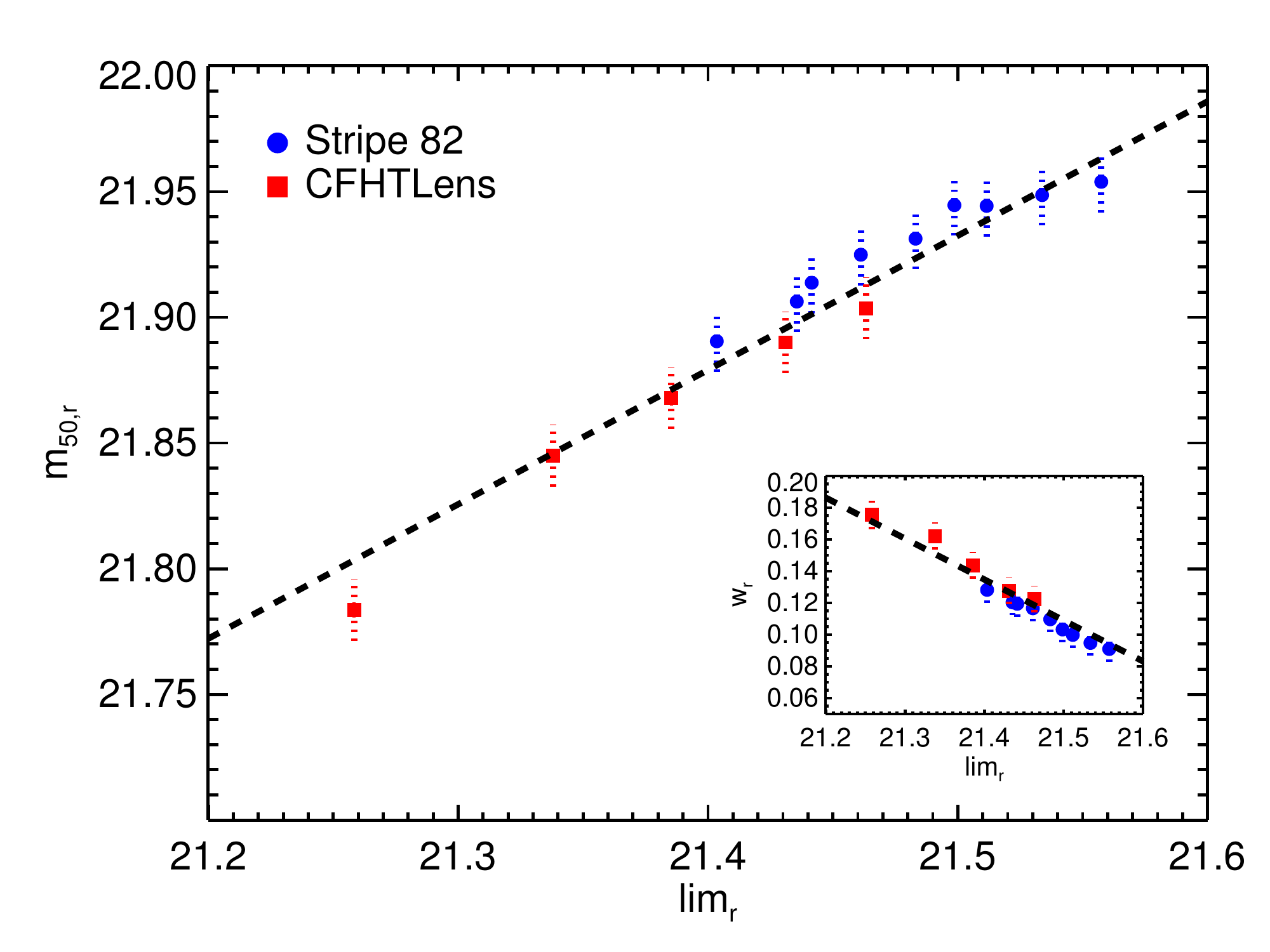}}
    \caption{\emph{Main panel:} Completeness parameter $m_{50}$, the magnitude at which $50\%$ of
      the galaxies are detected, as a function of local depth measured from
      $\rcmod$.  Blue circles are from S82, red squares from CFHTLens.  Black
      dashed line is a linear fit.  \emph{Inset}:Width parameter $w$ as a
      function of local depth.  Symbols are the same as the main panel.}
    \label{fig:m50r}
  \end{center}
\end{figure}

\begin{figure}
  \begin{center}
    \scalebox{1.0}{\plotone{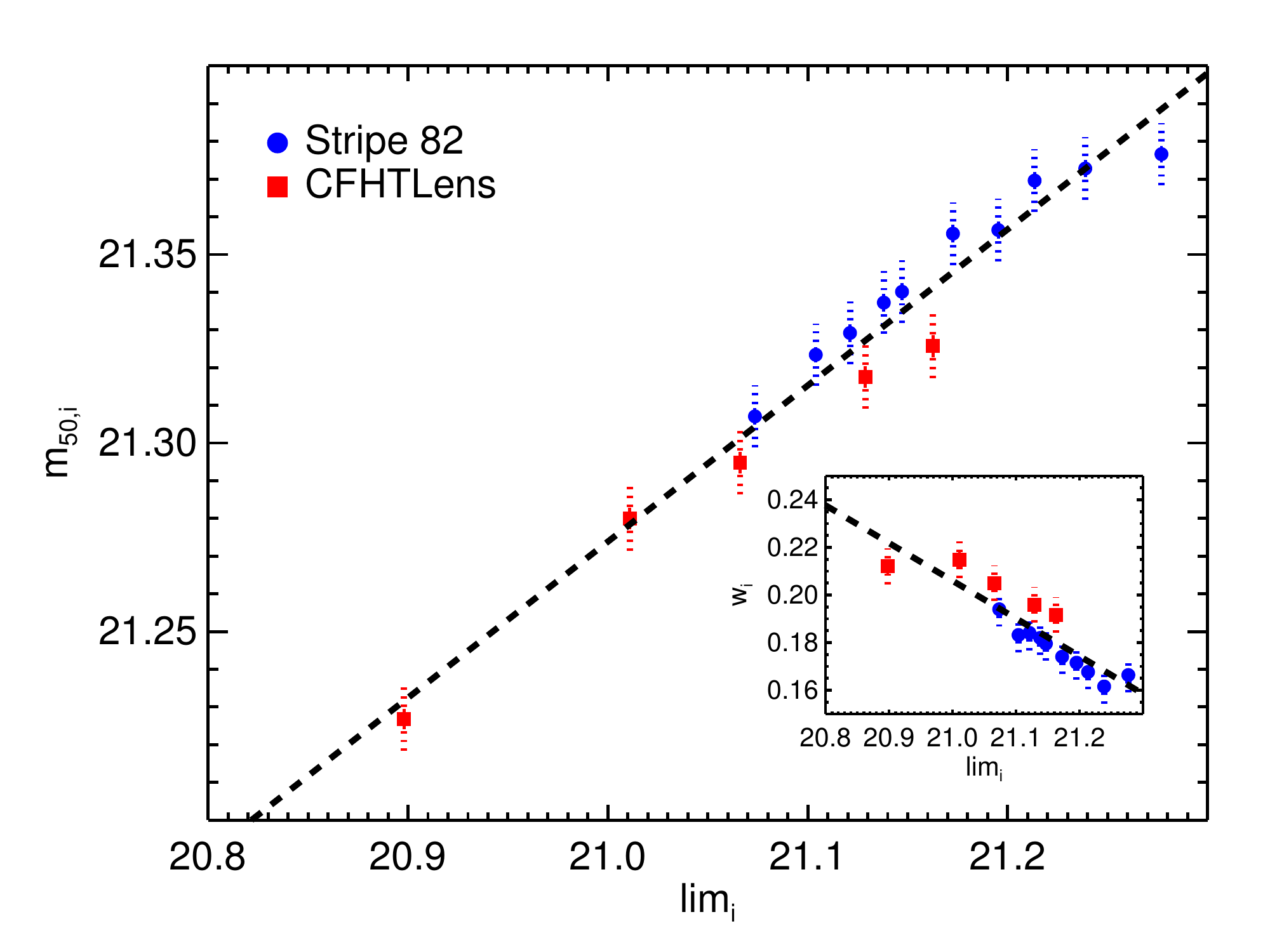}}
    \caption{Same as Figure~\ref{fig:m50r}, with limiting magnitude measured
      from $\icmod$.}
    \label{fig:m50i}
  \end{center}
\end{figure}

With these measurements in hand we determine how the galaxy completeness
parameter $m_{50}$ depends on the survey depth parameter $\mlim$.  
Figures~\ref{fig:m50r} and \ref{fig:m50i} show
the completeness as a function of the local limiting magnitude measured for
$\rcmod$ and $\icmod$ respectively.  Blue circles show the results for S82
regions, and red squares for CFHTLens.  The S82 regions tend to be deeper than
average (as discussed in Section~\ref{sec:computedepth}), and thus we sample a wide range
of local depth with some overlap between the two data sets.  The two data sets
give consistent results, with a small systematic offset of $\sim0.01$ magnitude.  Solid
error bars are those estimated from the bootstrap resampling, and the dashed
error bars are an estimate of the systematic error required to yield
$\chi^2/\mathrm{dof} = 1$ for the linear fit.

Table~\ref{tab:comppar} shows the final completeness parameter fits for
$m_{50}$ and $w$.  We note that both $m_{50}$ and $w$ are strong functions of
the local limiting magnitude.  Thus, in order to estimate the completeness at
any point in the survey, we must first know the local depth.

\begin{deluxetable*}{ccccc}
\tablecaption{Completeness Fit Parameters}
\tablehead{
  \colhead{Magnitude} &
  \colhead{$a_{\mathrm{comp}}$} &
  \colhead{$b_{\mathrm{comp}}$} &
  \colhead{$a_w$} &
  \colhead{$b_w$}}
\startdata
$\rcmod$ & $21.67\pm0.02$ & $0.53\pm0.04$ & $0.24\pm0.01$ & $-0.26\pm0.03$ \\
$\icmod$ & $21.27\pm0.01$ & $0.41\pm0.02$ & $0.206\pm0.003$ & $-0.16\pm0.02$ \\
\enddata
\tablecomments{$m_{50} = a_{\mathrm{comp}} + b_{\mathrm{comp}}(\mathrm{lim}
  - 21)$ ; $w = a_w + b_w(\mathrm{lim} - 21)$}
\label{tab:comppar}
\end{deluxetable*}

\section{Impact on Correlation Functions}
\label{sec:corrfxn}

As an example of the use of these depth maps, we now investigate the impact on
the estimation of correlation functions.  The measurement of the correlation
function $w(\theta)$ requires a set of random points that describes the window
function of the observable~\citep{landyszalay93}.  To first order, these random
points can be sampled uniformly from the geometric mask of the survey. Although
this works well at brighter magnitudes, when pushing the limits of a
photometric survey this is certain to break down.  In this section, we measure
a simulated correlation function while comparing different methods of
generating random points to show the utility of our method at faint magnitudes.

\subsection{Generating a Simulated Survey}

Our first task is to generate a simulated DR8-like survey with known
systematics and known correlation function.  We follow the same general plan as
in Section~\ref{sec:estnoise}, using a Monte Carlo technique to simulate data
with uniform density (zero correlation).  Consequently, any correlations observed
in the data must necessarily be due to systematics being imprinted into the galaxy
catalog.

There are two primary differences between this simulation and that from
Section~\ref{sec:estnoise}.  The first is that we are building in the
completeness estimation from Section~\ref{sec:completeness}.  The second, is that we
sample galaxies by color as well as magnitude to enable color selection in
random point generation.  Our procedure is as follows:
\begin{enumerate}
  \item{Generate simulated galaxies uniformly with zero correlation and
    approximately the same density as DR8 with $\icmod<21$.}
  \item{Give each galaxy a true magnitude vector sampled from the (deeper) S82
    catalog.}
  \item{Estimate the completeness for each galaxy based on $\icmod$.  Remove
    galaxies randomly according to the completeness estimate.}
  \item{For each magnitude of each of the remaining galaxies, perturb the input
    magnitudes in accordance to the input survey depth map to yield a simulated survey as in Section~\ref{sec:estnoise}.}
\end{enumerate}

As before, given this simulated survey we recompute the depth maps, fit the
depth map, reconstruct a simulated high resolution depth map, and estimate
the completeness from the simulated data.  It is these re-estimated maps that
are used in the following tests.

\subsection{Generating Random Points}

In order to measure the correlation function in real-space, we make use of the
{\tt CORR2} code~\citep{jbj04} for computing the Landy-Szalay statistic:
\be
\Omega = \frac{DD - 2DR + RR}{RR}.
\ee

For our measurements, $D$ refers to the simulated galaxies, and $R$ refers to
the random points used to sample the survey mask.  We have chosen to compare
the impact of four different methods of generating random points.  These are:
\begin{enumerate}
  \item{``Uniform'': Only the geometric mask has been applied.}
  \item{``Low Resolution'': The low resolution \emph{measured} depth maps
    (NSIDE=256) are used.  We also apply the completeness correction.}
  \item{``Reconstructed'': The high resolution \emph{reconstructed} depth maps
    (NSIDE=2048) are used. We also apply the completeness correction.}
  \item{``PSFMAG'': The $5\sigma$ point-source psf magnitude depths from
    Section~\ref{sec:psfmag} are used, along with the derived constant offsets
    so the values match the $10\sigma$ galaxy depths on average.  We also apply the completeness correction.}
\end{enumerate}

In all cases, the generation of random points follows the method for building
the simulated survey galaxies in the first place.  The primary difference is
that we use the \emph{new} reconstructed depth maps and re-estimated
completeness.  Therefore, the difference between the simulated survey galaxies and the random
points is the noise in the estimation of the depth and completeness maps.  

\subsection{Results}

In the interest of simplicity, we measure the correlation function for two
magnitude cuts in $i$-band, and ignore the impact of the color selection.  The
first magnitude cut estimates the impact of depth variations on the correlation
function for the faintest BOSS CMASS
galaxies~\citep[$19.8<\icmod<20.0$;][]{wbbst11,rhctp11,hcsdr12}.  The second
magnitude cut estimates the impact of depth variations on photometrically
detected galaxies near the average $10\sigma$ survey limit, $20.9<\icmod<21.0$.

\begin{figure}
  \begin{center}
    \scalebox{1.2}{\plotone{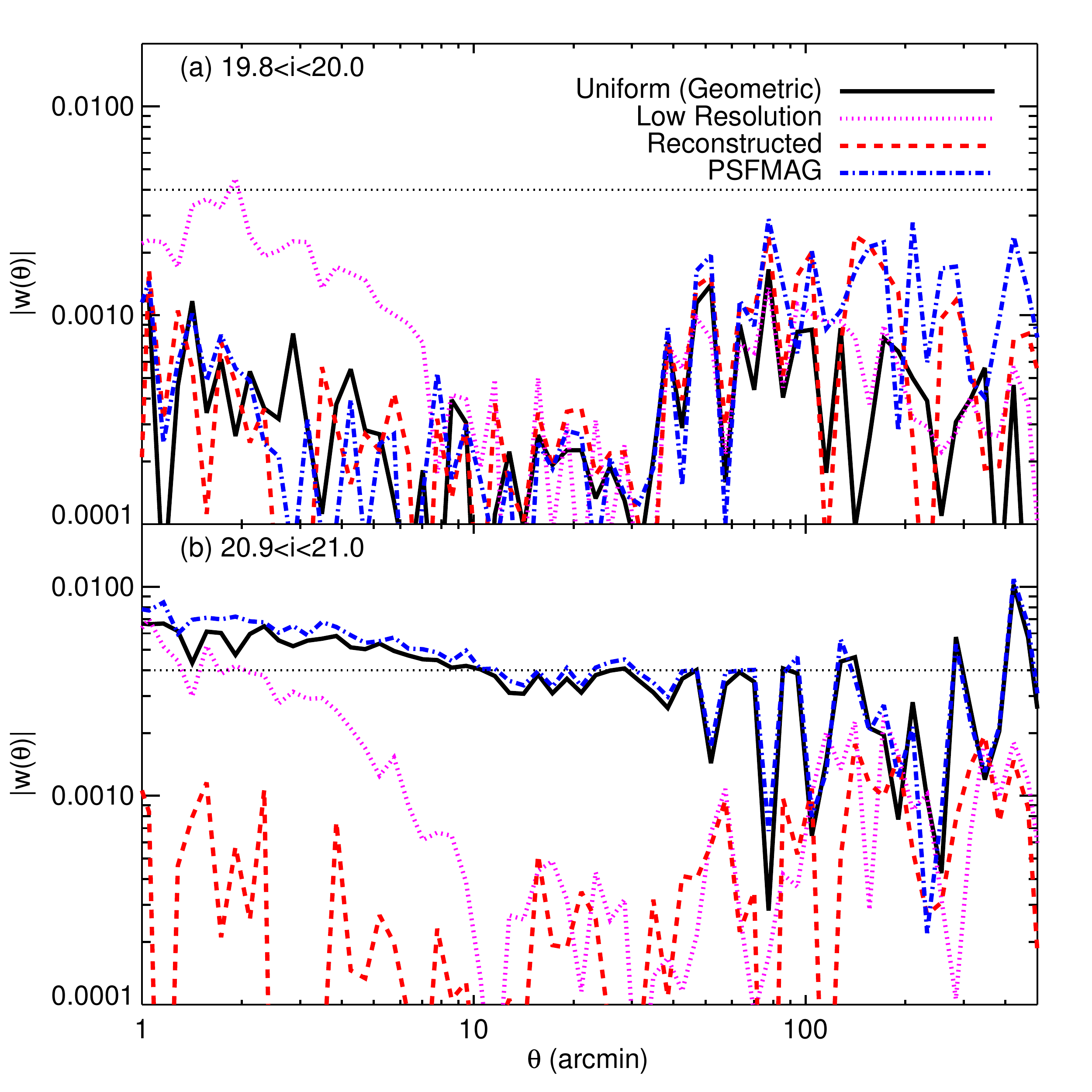}}
    \caption{Absolute value of the correlation function $|w(\theta)|$ for our simulated
    survey data using different sets of random points.
      \emph{Top Panel:} Magnitude selection equivalent to the faintest BOSS
      CMASS galaxies, $19.8<\icmod<20.0$.  The black dotted line shows the
      typical error on $w(\theta)$ from \citet{hcsdr12}.  The solid black line
      shows the bias from using uniform randoms (with the geometric mask).  The
      magenta dotted line shows the bias from the ``Low Resolution'' randoms.
      The red dashed line shows the bias from the full ``Reconstructed''
      randoms, and the blue dot-dashed line shows the bias from the ``PSFMAG''
      randoms.    \emph{Bottom Panel:} Same as above, for
      galaxies near the $10\sigma$ survey limit, $20.9<\icmod<21.0$.  The
      impact on $w(\theta)$ for uniform randoms is large at all scales.  Only
      the full reconstructed randoms show nearly unbiased results at all scales.
    }
    \label{fig:correlation}
  \end{center}
\end{figure}

Figure~\ref{fig:correlation} shows the absolute value of the two-point
correlation function $|w(\theta)|$ estimated for various methods of generating
random points.  The dotted black line shows the typical error on $w(\theta)$
from \citet{hcsdr12}.  Any additive bias on the correlation function below this
level is sub-dominant.  In the top panel, for the faintest CMASS galaxies with
$19.8<\icmod<20.0$, we can see that uniform randoms (sold black line) are adequate
to compute $w(\theta)$.  Similarly, the other random point generation schemes,
with the same geometric mask, all perform adequately.  The only possible
exception is the ``Low Resolution'' method (magenta dotted line) which may
impart some structure at very small scales.  In the bottom panel, for the
galaxies near the $10\sigma$ photometric limit of the survey ($20.9<\icmod<21.0$),
the choice of random points is much more significant.  Using uniform random
points (with the geometric mask) leads to significant residuals at all scales.
The ``PSFMAG'' estimate of the depth (blue dot-dashed line) is better than
uniform, but still has problems at all scales.  The ``Low Resolution'' method
works at large scales but not at small scales, while our full ``Reconstructed''
method (red dashed line) yields the best results, with only a small bias
at the smallest scales.  Therefore, we are confident that this method
accurately captures depth and completeness variations down to faint magnitudes.

\section{Results and Summary}
\label{sec:summary}

In this paper, we have introduced a new way of empirically constructing depth
maps from large photometric survey data.  By combining low resolution depth
maps with high resolution maps of survey systematics, we can use machine
learning to reconstruct high resolution depth maps for typical galaxies
measured using any arbitrary magnitude definition.  Using SDSS DR8 as an
example, our method can reconstruct the depth at an arbitrary point in the
survey to within $\sim2\%$, thus making it possible to accurately measure
correlation functions for galaxies detected near the limiting magnitude of the
survey.  We emphasize that this method requires the knowledge of both a precise
geometric mask, as well as accurate maps of the various systematics that affect
galaxy photometry, especially seeing and sky brightness.  Furthermore, the reported
photometric errors in the data must be correct: any biases in the estimated
photometric uncertainties of the detected galaxies will propagate into the survey
depth maps.

In order to accurately build sets of random points suitable for correlation
function measurements, a depth map is insufficient.  We must also estimate the
completeness function for galaxy detection.  To estimate this function directly
from the data, it is necessary to have data that is significantly deeper than
the main survey.  Fortunately, this is not only available for SDSS in the
Stripe 82 coadds, but it is a common aspect of current and future wide-field
surveys, including the Medium-Deep Survey fields in
Pan-STARRs1~\citep{panstarrs02} and the supernova search fields in
DES~\citep{des05}.  We have shown that the completeness can be parametrized by
a simple function for SDSS, and that the parameters of this function depend on
the local survey depth.  In this way, the depth map reconstruction is an
essential prerequisite for estimating the local completeness of the survey.

As a sample of the utility of these depth maps, we demonstrate a new way of
building random points for correlation function measurements that incorporates
the full knowledge of the survey depth and local completeness.  Uniform random
points are sufficient when one is far from the photometric limit of the
survey, as is the case for BOSS CMASS galaxies.  These galaxies are relatively
unaffected by variations in the local depth for two main reasons.  First, the
drift-scan technique ensures that SDSS is a relatively uniform survey, without
large variations in depth.  Second, we note that while SDSS uses the same
telescope for both photometric and spectroscopic measurements, by necessity any
spectroscopic target selection (such as CMASS) will necessarily be
significantly brighter than the photometric limit for any reasonable balance of
exposure time between the two surveys.  Thus, it is not surprising that for the
CMASS sample, designed for spectroscopic follow-up, uniform random points (with
a proper geometric mask) are sufficient.

However, for ongoing and future photometric surveys, this will not be the case.
In particular, both DES~\citep{des05} and LSST~\citep{lsst08} are purely
photometric surveys where the best science will not be achieved by limiting
measurements to galaxies more than a magnitude brighter than the survey limit.
Furthermore, with tiled observations and chip gaps, depth variations will be
much more significant than in SDSS, especially in the early phases before
complete coverage is achieved. In addition, surveys such as the DECam Legacy
Survey~(DECaLS\footnote{http://legacysurvey.org}) will be used for target
selection for the Dark Energy Spectrographic Instrument
Survey~(DESI\footnote{http://desi.lbl.gov}) are not significantly deeper than
the spectroscopic targeting. Therefore, we expect this method of building depth
maps will be useful beyond the current application to SDSS.

\acknowledgements

This work was supported in part by the U.S. Department of Energy contract to
SLAC no. DE-AC02-76SF00515.

Funding for SDSS-III has been provided by the Alfred P. Sloan Foundation, the Participating Institutions, the National Science Foundation, and the U.S. Department of Energy Office of Science. The SDSS-III web site is http://www.sdss3.org/.

SDSS-III is managed by the Astrophysical Research Consortium for the Participating Institutions of the SDSS-III Collaboration including the University of Arizona, the Brazilian Participation Group, Brookhaven National Laboratory, University of Cambridge, Carnegie Mellon University, University of Florida, the French Participation Group, the German Participation Group, Harvard University, the Instituto de Astrofisica de Canarias, the Michigan State/Notre Dame/JINA Participation Group, Johns Hopkins University, Lawrence Berkeley National Laboratory, Max Planck Institute for Astrophysics, Max Planck Institute for Extraterrestrial Physics, New Mexico State University, New York University, Ohio State University, Pennsylvania State University, University of Portsmouth, Princeton University, the Spanish Participation Group, University of Tokyo, University of Utah, Vanderbilt University, University of Virginia, University of Washington, and Yale University.

\appendix

\section{Reconstructed Depth Maps}

The maps are available at {\tt
  http://risa.stanford.edu/redmapper}.  The list of maps are in
Table~\ref{tab:mapnames}.  Each of them is in \healpix{} FITS format, NSIDE=2048,
ring ordered, equatorial coordinates.

\begin{deluxetable*}{ccl}
\tablewidth{0pt}
\tablecaption{List of \healpix{} Maps}
\tablehead{
  \colhead{Band} &
  \colhead{Type} &
  \colhead{Name}
}
\startdata
$u$ & model & {\tt sdss\_dr8\_nodered\_nside2048\_u\_model\_10sigma.fits.gz} \\
$g$ & model & {\tt sdss\_dr8\_nodered\_nside2048\_g\_model\_10sigma.fits.gz} \\
$r$ & model & {\tt sdss\_dr8\_nodered\_nside2048\_r\_model\_10sigma.fits.gz} \\
$i$ & model & {\tt sdss\_dr8\_nodered\_nside2048\_i\_model\_10sigma.fits.gz} \\
$z$ & model & {\tt sdss\_dr8\_nodered\_nside2048\_z\_model\_10sigma.fits.gz} \\
$r$ & cmodel & {\tt sdss\_dr8\_nodered\_nside2048\_r\_cmodel\_10sigma.fits.gz} \\
$i$ & cmodel & {\tt sdss\_dr8\_nodered\_nside2048\_i\_cmodel\_10sigma.fits.gz} \\
& $E(B-V)$ & {\tt sdss\_dr8\_ebv\_sfd98\_nside2048.fits.gz} \\

\enddata
\tablecomments{All files are in \healpix{} FITS format, NSIDE=2048, ring
  ordered, equatorial coordinates.}
\label{tab:mapnames}
\end{deluxetable*}

\end{document}